\documentclass{aastex62}
\usepackage{booktabs}
\usepackage{float}
\usepackage{changepage}
\usepackage[graphicx]{realboxes}
\usepackage{CJK}
\def\lea{\mathrel{<\kern-1.0em\lower0.9ex\hbox{$\sim$}}}
\def\gea{\mathrel{>\kern-1.0em\lower0.9ex\hbox{$\sim$}}}

\graphicspath{{./}{}}
\received{}
\revised{}
\accepted{}
\submitjournal{ApJ}

\shorttitle{}
\shortauthors{}

\def\aap{A\&A}

\def\apjl{ApJ}

\def\apjs{ApJS}

\def\lea{\mathrel{<\kern-1.0em\lower0.9ex\hbox{$\sim$}}}
\def\gea{\mathrel{>\kern-1.0em\lower0.9ex\hbox{$\sim$}}}

\newcommand{\OSU}{\affiliation{Department of Astronomy, The Ohio State University, 140 West 18th Avenue, Columbus, Ohio 43210, USA}}

\newcommand{\Alberta}{\affiliation{Department of Physics, University of Alberta, Edmonton, AB T6G 2E1, Canada}}

\newcommand{\ANU}{\affiliation{Research School of Astronomy and Astrophysics, Australian National University, Canberra, ACT 2611, Australia}}

\newcommand{\ESO}{\affiliation{European Southern Observatory, Karl-Schwarzschild Stra{\ss}e 2, D-85748 Garching bei M\"{u}nchen, Germany}}

\newcommand{\Heidelberg}{\affiliation{Astronomisches Rechen-Institut, Zentrum f\"{u}r Astronomie der Universit\"{a}t Heidelberg, M\"{o}nchhofstra\ss e 12-14, D-69120 Heidelberg, Germany}}

\newcommand{\JHU}{\affiliation{Department of Physics and Astronomy, The Johns Hopkins University, Baltimore, MD 21218, USA}}

\newcommand{\MPIA}{\affiliation{Max-Planck-Institut f\"{u}r Astronomie, K\"{o}nigstuhl 17, D-69117, Heidelberg, Germany}}

\newcommand{\OCAD}{\affiliation{OCAD University, Toronto, Ontario, M5T 1W1, Canada}}

\newcommand{\UToledo}{\affiliation{Ritter Astrophysical Research Center, University of Toledo, Toledo, OH, 43606}}

\newcommand{\ULyon}{\affiliation{Univ Lyon, Univ Lyon 1, ENS de Lyon, CNRS, Centre de Recherche Astrophysique de Lyon UMR5574,\\ F-69230 Saint-Genis-Laval, France}}

\newcommand{\UMichigan}{\affiliation{University of Michigan}}

\newcommand{\UniCA}{\affiliation{Université Côte d'Azur, Observatoire de la Côte d'Azur, CNRS, Laboratoire Lagrange, 06000, Nice, France}}

\newcommand{\UWyoming}{\affiliation{Department of Physics and Astronomy, University of Wyoming, Laramie, WY 82071, USA}}

\newcommand{\STScI}{\affiliation{Space Telescope Science Institute, 3700 San Martin Drive, Baltimore, MD 21218, USA}}

\newcommand{\Tenerife}{\affiliation{Instituto de Astrof\'isica de Canarias, calle Vía L\'actea s/n, E-38205 La Laguna, Tenerife, Spain}}
\newcommand{\Laguna}{\affiliation{Departamento de Astrof\'isica, Universidad de La Laguna, Avenida Astrof\'isico Francisco S\'anchez s/n, E-38206 La Laguna, Spain}}

\newcommand{\Ensenada}{\affiliation{Instituto de Astronom\'ia, Universidad Nacional Aut\'onoma de M\'exico, Unidad Acad\'emica en Ensenada, Km 103 Carr. Tijuana$-$Ensenada, Ensenada, B.C.,\\C.P. 22860, M\'exico}}

\newcommand{\Cambridge}{\affiliation{Institute of Astronomy and Kavli Institute for Cosmology, University of Cambridge, Madingley Road, Cambridge, CB3 0HA, UK}}

\begin{document}
\title{Star Cluster Populations in 38 Spiral Galaxies: Evidence for Near-Universal Mass and Age Distributions from PHANGS}


\author[0000-0003-0085-4623]{Rupali~Chandar}\email{Rupali.Chandar@utoledo.edu}
\UToledo
\author[0009-0008-2929-6665]{Matthew~R.~Floyd}
\UToledo
\author[0000-0002-8528-7340]{David~A.~Thilker}
\JHU
\author[0000-0001-7448-1749]{Kiana Henny}
\UWyoming
\author[0000-0002-3784-7032]{Bradley C. Whitmore}
\STScI
\author[0000-0002-2278-9407]{Janice C. Lee}
\STScI
\UniCA
\author[0000-0003-1943-723X]{Sinan Deger}
\Cambridge
\author[0000-0002-5782-9093]{Daniel A. Dale}
\UWyoming
\author[0000-0003-3917-6460]{Kirsten L. Larson}
\STScI
\author[0000-0003-0946-6176]{Médéric Boquien}
\UniCA
\author[0000-0001-9852-9954]{Oleg Gnedin}
\UMichigan
\author[0000-0001-6038-9511]{Daniel Maschmann}
\STScI
\author[0000-0002-2545-1700]{Adam K. Leroy}
\OSU
\author[0000-0002-5204-2259]{Erik~Rosolowsky}
\Alberta
\author[0000-0001-7130-2880]{Leonardo \'Ubeda}
\STScI
\author[0000-0003-0410-4504]{Ashley~T.~Barnes}
\ESO
\author[0000-0002-3933-7677]{Eva Schinnerer}
\MPIA
\author[0000-0002-6155-7166]{Eric Emsellem}
\ESO
\ULyon
\author[0000-0001-7413-7534]{Angus Mok}
\OCAD
\author[0000-0002-3247-5321]{Kathryn~Grasha}
\altaffiliation{ARC DECRA Fellow}
\ANU 
\author[0000-0001-5965-3530]{Francesca Pinna} 
\Tenerife
\Laguna
\author[0000-0001-8289-3428]{Aida Wofford}
\Ensenada
\author[0000-0002-0560-3172]{Ralf S.\ Klessen}
\Heidelberg
\author[0000-0001-6708-1317]{Simon Glover}
\Heidelberg
\author[0000-0002-4781-7291]{Sumit K. Sarbadhicary}
\JHU
\author[0000-0001-7113-8152]{Ivan Gerasimov}
\UniCA

\begin{abstract}
We present the largest study to date of star cluster mass and age distributions in nearby galaxies. The analysis is based on a uniformly selected catalog of over 15,000 (36,000) human (machine) classified compact clusters across 38 spiral galaxies in the PHANGS-HST survey, with improved photometric SED age dating that mitigates key degeneracies with reddening and metallicity. The galaxies span a factor of $\sim$100 in star formation rate (SFR) and in their surface density of star formation and molecular gas ($\Sigma_{\rm SFR}$, $\Sigma_{\rm H_2}$), and cover a broad range of morphologies. We find that cluster mass functions are well described by a power law, $dN/dM\propto M^{\beta}$, with $\beta=-1.9\pm0.2$, and that this shape is remarkably similar across all galaxies in our sample and eight composite groups of galaxies with similar SFRs, with no significant dependence on SFR, $\Sigma_{\rm SFR}$, or $\Sigma_{\rm H_2}$.
Cluster age distributions decline steeply and can be described by $dN/d\tau\propto\tau^{\gamma}$ with $\gamma=-0.83\pm0.24$, again with little dependence on host-galaxy properties or cluster mass. The independence of the mass and age distributions implies a separable joint distribution, $g(M,\tau)\propto M^{\beta}\tau^{\gamma}$, and suggests that cluster populations in spiral disks are dominated by strong, mass-independent dissolution over their first $\sim$Gyr. These results establish a benchmark for interpreting cluster demographics across diverse galactic environments and provide context for studies of young cluster populations at high redshift.
\end{abstract} 
\section{Introduction}\label{sec-introduction}

Stellar clusters are a fundamental unit of star formation, and powerful tracers of the physical processes that regulate how stars form, interact with their environment, and populate galaxies. Clusters are born in the densest, dust-enshrouded regions of molecular clouds \citep{lada03}, and rapidly emerge on timescales of a few Myr as feedback from massive stars clears their natal gas through ionization, radiation pressure, stellar winds, and supernovae \citep{hollyhead15,hannon19,hannon22,ostriker22,Rodriguez23,Rodriguez25,evaadam24}. From birth onward, clusters steadily lose mass and many ultimately dissolve, releasing their stars into the surrounding field. The imprint of these formative and disruptive processes is encoded in the observed mass and age distributions of cluster populations, making cluster demographics a uniquely sensitive probe of star formation, feedback, and dynamical evolution within galaxies \citep[e.g.,][and references therein]{Krumholz19}.

Most star and cluster formation today takes place in gas-rich spiral disks whose global star formation histories (SFHs) have evolved relatively smoothly over the past $\sim$Gyr.
In particular, nearby spiral galaxies in the Local Volume generally exhibit fairly steady SFHs on galactic scales, despite larger fluctuations on smaller spatial and temporal scales.  Resolved stellar population studies of systems such as M31, M33, and NGC~7793 find variations in their global star formation rates of only factors of $\sim$1.5 over the past Gyr \citep[e.g.,][]{Lewis15,Lazzarini22,Sacchi18,SilvaVilla11}, while comparisons between H$\alpha$ and far-ultraviolet luminosities similarly indicate that many nearby star-forming galaxies have maintained broadly stable recent star formation activity \citep{Lee09} over the past few hundred Myr. Although localized regions within disks can experience substantial variations in star formation on $\sim100$~pc scales, these fluctuations largely average out over several-kpc scales. Spiral galaxies therefore provide favorable laboratories for studying cluster evolution, since changes in cluster demographics are less strongly coupled to large variations in the underlying cluster formation rate.

A number of theoretical works have predicted that lower-mass clusters dissolve faster than higher-mass ones in spiral disks, 
i.e.\ that dissolution is mass-dependent. \citet{Lamers05} developed a simple analytic model for mass-dependent dissolution, where the characteristic disruption time t$_{\rm dis}$ scales with the initial cluster mass as t$_{\rm dis} \propto M^{0.62}$. In this model t$_{\rm dis}$ also depends on the density of the disk environment, so clusters in higher (lower) density environments (characterized by the surface density of star formation $\Sigma_{\rm SFR}$ or molecular gas $\Sigma_{\rm H_2}$) dissolve faster (slower). The disruption {\em rate} is encoded in the steepness of the cluster age distribution, with a low dissolution rate producing a flat distribution and strong dissolution producing a steep one when star formation is constant. Mass-dependent dissolution results in the coupling between mass and age distributions for an evolving cluster population, with a steeper age distribution for lower-mass clusters and a flatter mass function for older clusters. \citet{Kruijssen11} combined an N-body/smooth particle hydrodynamics galaxy evolution code with the semi-analytic dissolution models of \citet{Lamers05}, included several dissolution mechanisms (stellar feedback, stellar evolution, two-body relaxation, and tidal shocks) and used the simulations to predict the evolution of cluster populations. They found that the mean dissolution rate decreases as clusters age and escape their dense birth environment, and predicted that the cluster age distribution should be steeper in spiral galaxies with higher $\Sigma_{\rm SFR}$.  

Other works have suggested that early cluster evolution, before two-body relaxation begins to erode the population, is mostly independent of cluster mass, meaning that clusters with different masses experience similar dissolution times \citep[e.g.,][]{fall05,whitmore07,chandar10a,fall12}. Mass-independent dissolution is expected to drive cluster demographics for tidally-limited clusters since their density is not correlated with mass and the dissolution timescale depends on cluster properties only through their density \citep[e.g.,][]{Binney08,fall12}. The shapes of cluster mass and age distributions are independent of one another when mass-independent dissolution dominates the demographics. A higher rate of dissolution will result in a steeper age distribution and larger drop in the amplitude of the cluster mass function for older clusters \citep[e.g.,][]{whitmore07,fall09}.

Observationally, cluster age-mass demographics have been analyzed in fewer than a dozen nearby spiral galaxies,
with most studies focusing on a single galaxy. As summarized by \citet{Krumholz19}, these empirical studies have not yet converged, often finding conflicting results for the same galaxy (sometimes from the same observations). For the age distribution, findings for the power-law index $\gamma$ range from nearly flat with $\gamma\approx-0.2$ (i.e., little dissolution) in M31 \citep{Fouesneau14}, M83 \citep[][]{SilvaVilla11}, NGC628 \citep{adamo17}, NGC~7793 \citep{SilvaVilla11}, and NGC~2997 \citep{ryon14} to quite steep with $\gamma$ between $\approx-0.7$ and $-1.0$ (i.e. strong dissolution) for M83 \citep{chandar10c}, M51 \citep{chandar16}, NGC~628 \citep{Whitmore23} over the past few$\times100$~Myr. Some works find evidence that the age distribution is steeper for lower-mass clusters as expected for mass-dependent dissolution (e.g., in NGC 628, M101, see \citealt{adamo17,Linden22}, respectively) while others find similar shapes for clusters with different masses as expected from mass-independent dissolution \citep[e.g.,][]{chandar10c,chandar16,Whitmore23}. There have also been suggestions that both mass-dependent and mass-independent dissolution drive cluster demographics in some spiral galaxies \citep[e.g.,][]{bastian12a,adamo17}.

Three main issues have likely contributed to the conflicting observational results: (1) cluster selection, (2) the age-dating process, and (3) various assumptions made when constructing the empirical distributions. For selection, some studies restrict their sample to clusters that appear morphologically compact or dynamically concentrated in an attempt to distinguish bound clusters from unbound associations \citep[e.g.,][]{bastian11,bastian12a,adamo15}. Such morphological or dynamical criteria do not necessarily provide a reliable indication of a gravitationally bound cluster, so other studies adopt a broader definition \citep[e.g.,][]{Chandar2010,chandar11,chandar14}. Different selection criteria used between different works primarily affect clusters younger than $\sim$10~Myr, with differences of $\approx30-50$\% in the number of very young clusters \citep{Krumholz19}. Uncertainties related to selection criteria can be addressed by including fits with and without clusters younger than $\lea10$~Myr.

The specific method and assumptions used to age-date clusters can have a surprising impact on the resulting age-mass distributions. Disentangling the key degeneracies between age and reddening for young clusters, and age and metallicity for ancient ones are challenging when only broad-band photometry is used for spectral energy distribution (SED) fitting, even when measurements in the U and NUV filters are included \citep[e.g.,][]{Whitmore99,hannon19,turner21,deger22,hannon22,Whitmore23}. \citet{Whitmore23} found that challenges in breaking the age-reddening degeneracy likely result in an artificial steepening of the age distribution of lower-mass relative to higher-mass clusters in NGC~628 \citep{adamo17}, and that the difference in shape disappeared when an improved method of breaking the age-reddening degeneracy was instead applied. Ancient globular clusters can also suffer from poor age-dating results in standard SED fitting, as most large surveys assume a solar metallicity simple stellar population model for clusters at all ages \citep[e.g.,][]{adamo17}. \citet{Whitmore23} and \citet{Floyd24} found that more than 80\% of globular clusters were incorrectly dated to ages younger than 1~Gyr in the initial PHANGS-HST cluster pipeline \citep[][]{turner21}. 

The main goal of this work is to establish the empirical shapes of star cluster mass and age distributions in nearby spiral galaxies across a broad range of galactic environments. Previous observational studies have reached conflicting conclusions regarding the strength, mass dependence, and environmental dependence of cluster dissolution, due in part to heterogeneous cluster selection, uncertainties in age-dating, and limitations in the size and diversity of galaxy samples. The PHANGS-HST survey now provides the largest homogeneous sample available for a systematic comparison of cluster demographics across a representative sample of nearby spiral galaxies \citep{Lee2022}. Our primary goals are to determine: (1) whether cluster dissolution rates in nearby spiral galaxies are generally weak or strong, (2) whether dissolution timescales depend on cluster mass, and (3) if there are significant variations in dissolution timescales that correlate with the star-forming properties (characterized here by $\Sigma_{\rm SFR}$ and $\Sigma_{\rm H_2}$) of the host galaxy. We also examine composite cluster populations formed by combining galaxies with similar global star formation rates, allowing us to improve statistical power and reduce the impact of possible variations in recent star formation histories. 

We use the PHANGS-HST star cluster catalogs \citep{maschmann24,Thilker25}, containing nearly $\sim$100,000 star clusters and compact stellar associations across 38 nearby spiral galaxies. The sample spans factors of $\approx100$ in SFR, $\Sigma_{\rm SFR}$, and $\Sigma_{\rm H_2}$, enabling a systematic investigation of how cluster demographics vary with galactic environment \citep{Lee2022}. 
As summarized in \citep{Lee2022} the PHANGS-HST cluster catalogs were constructed by the PHANGS Stellar Populations Working Group using a uniform selection and analysis pipeline, and incorporate improved photometry, morphological classification, host galaxy distance estimates, as well as age and mass estimates designed to mitigate key degeneracies between age, reddening, and metallicity \citep{whitmore21, deger22, Anand21, Thilker22, wei20, hannon23}. Throughout this work we compare results derived from both individual galaxies and composite galaxy groups, as well as from human- and machine-learning-selected cluster catalogs and two independent age-dating approaches. The consistency between these different analyses demonstrates that the main empirical trends are robust to the these particular choices.

The rest of this paper is organized as follows. Section~\ref{sec-sample} presents the galaxy sample and defines eight composite groups, each containing 3–5 galaxies with similar global star formation rates. As mentioned earlier, the cluster populations within each group are combined to improve the sampling statistics throughout the analysis. Section~\ref{sec-sample} also describes the cluster catalogs, summarizes their selection and age-dating procedure, and assesses incompleteness. Section~\ref{sec-Mt} reports on the resulting age-mass diagrams, while Sections~\ref{sec-cmf} and \ref{sec-dndt} quantify the empirical cluster mass and age distributions for the eight galaxy subsets and for individual galaxies. We discuss the implications of our observational results for cluster dissolution in Section~\ref{sec-discussion}. Our main conclusions and potential future work are summarized in Section~\ref{sec-conclusions}. An Appendix presents a additional results for the cluster mass and age distributions of individual galaxies.

\section{Galaxy Sample and Star Cluster Catalogs}\label{sec-sample}

\subsection{Spiral Galaxies and Star-forming Groups}

The PHANGS (Physics at High Angular resolution in Nearby GalaxieS) collaboration has collected a rich, multi-wavelength dataset for $\sim$90 nearby spiral galaxies\footnote{\url{https://phangs.stsci.edu}}. In this work, we focus on the cluster populations in a subset of 38 galaxies that were observed as part of cycle 26 HST Treasury program 15654 (PI: J. Lee, \citealt{Lee2022}), optimally selected for the UV-optical study of resolved stellar populations (i.e., low inclination, high galactic latitude, and absence of bright foreground stars). The sample galaxies are located at distances between $\approx5-20$~Mpc \citep{Anand21}, have stellar masses between $5\times10^9-10^{11}~M_{\odot}$ \citep{Leroy21}, and include most spiral morphologies, such as barred and unbarred, grand-design and flocculent, with types from Sa through Sd \citep[][Table 1 and Figure 1]{Lee2022}. The total star formation rate (SFR$_{\rm tot}$) and that within the HST footprint of each galaxy (SFR) were determined from the far-ultraviolet flux measured from GALEX images plus the infrared luminosity measured from WISE W4 images, following the prescription given in \citet{Leroy19}. The SFR within the HST footprint (SFR) is one of the primary parameters considered in this work, and the galaxies are organized from highest to lowest SFR in Column~3 of Table~1. The sample covers a factor of $\approx100$ in SFR, from 12.9$~M_{\odot}~\mbox{yr}^{-1}$ in NGC~1365 to 0.1$~M_{\odot}~\mbox{yr}^{-1}$ in IC~5332. 
The SFR per unit area $\Sigma_{\rm SFR}$ compiled in Column~5 is also calculated within the HST footprint \citep[][]{Sun22,Sun2023}. 
The surface density of molecular gas $\Sigma_{\rm H_2}$ is taken from \citet[][]{Leroy21,Sun25} and compiled in column~6 of Table~\ref{tab:galaxysample}. Our sample galaxies also cover a range of $\approx100$ in $\Sigma_{\rm SFR}$ and $\Sigma_{\rm H_2}$.
\begin{deluxetable*}{lccccccccc}
\tablecaption{Spiral Galaxy Sample and Properties\label{tab:galaxysample}}
\centering
\tablehead{
\colhead{Galaxy} & \colhead{D} &  \colhead{SFR} & \colhead{SFR$_{\rm tot}$} &  \colhead{$\Sigma_{\rm SFR}$} & \colhead{$\Sigma_{\rm H_2}$}   & \colhead{\# Cl} & \colhead{\# Cl} & \colhead{Group}
 \\ 
\colhead{Name} & \colhead{(Mpc)} &
\colhead{($M_{\odot}~\mbox{yr}^{-1}$)}  & \colhead{($M_{\odot}~\mbox{yr}^{-1}$)} & \colhead{($M_{\odot}~\mbox{yr}^{-1}~\mbox{kpc}^{-2}$)} & \colhead{($M_{\odot}~\mbox{pc}^{-2}$)}  &  \colhead{Cat~1} & \colhead{Cat~2} & \colhead{\#}
\\
\colhead{(1)} & \colhead{(2)} & \colhead{(3)} & \colhead{(4)} & \colhead{(5)} & \colhead{(6)} & \colhead{(7)} & \colhead{(8)} & \colhead{(9)} 
}
\startdata
NGC~1365 & $19.6\pm 0.8$ & 12.9 &  16.9  & 0.242 & 232.18	   & 611 & 758 &  1 \\ 
NGC~1672 & $19.5\pm2.9$ & 6.6 & 7.6   & 0.081 & 62.63   & 372 & 2003 &  1 \\ 
NGC~4303 & $17.0\pm3.0$  & 4.3 &  5.4  & 0.035 & 52.68  & 556 & 1798 & 1\\ 
NGC~1097 & $13.6\pm2.0$ & 3.9  & 4.7 & 0.139 & 118.60   & 613 & 1725 & 1\\
NGC~1559 & $19.4\pm0.4$ & 3.8 & 3.8 & 0.025 & 15.71    & 709 & 1438 & 1 \\ \hline
NGC~3627     & $11.3\pm0.5$  & 3.3 & 3.9  &  0.029 & 43.96  & 773  & 2683 &  2 \\ 
NGC~1792 & $16.2\pm2.4$ & 3.2 & 3.7 &  0.177 & 28.52    & 563 & 728 & 2 \\
NGC~1566        & $17.7\pm2.0$  & 3.2  & 4.6  & 0.030 & 46.97  & 684 &  1271  & 2\\ 
NGC~4536 & $16.25\pm1.1$ & 3.1 & 3.5  & 0.016 & 11.99     & 313 & 716 & 2\\
NGC~4654 & $22.0\pm1.2$ & 3.1 & 3.8  & 0.0097 & 14.77     & 610 & 629 & 2 \\ \hline
NGC~4254      & $13.1\pm2.8$  & 2.8  & 3.1 & 0.049 & 90.10    & 480 & 2166 & ... \\ 
NGC~4321      & $15.2\pm0.5$   & 2.4 & 3.5 & 0.0136	& 27.08  & 712   & 1451  & 3 \\ 
NGC~2903 & $9.6\pm0.4$ & 2.4 & 3.1 & 0.018 & 21.54  & 499 &  1658 & 3 \\
NGC~1385       & $17.2\pm2.6$   & 2.0   & 2.1   & 0.031 & 21.49  & 521  & 535 & 3 \\
NGC~7496  & $18.7\pm2.8$  & 2.0 & 2.2   & 0.037 & 19.39  & 261  & 280 & 3 \\ \hline
NGC~5248 & $14.9\pm1.3$ & 1.7 & 2.3 & 0.020 & 37.19   & 526 & 709 & ... &  \\
NGC~1087  & $15.9\pm2.2$  & 1.3 & 1.3 & 0.0195 & 17.96  & 504 & 557 & 4 \\ 
NGC~4569 & $15.8\pm2.4$ & 1.2 & 1.3 & 0.0037 & 10.01 & 414 & 486 & 4 \\
NGC~4535 & $15.8\pm0.4$ & 1.2 & 2.2 & 0.0067 & 11.79   & 401 & 499 & 4 \\ 
NGC~1300 & $19.0\pm2.9$ & 1.0  & 1.2  & 0.00205 & 6.24 & 304 & 1463 &   4 \\ \hline
NGC~628 & $12.8\pm2.1$ & 0.9 & 1.8  & 0.0085 & 15.17  & 481 & 1662 & 5 \\
NGC~3351 & $10.0\pm0.3$ & 0.9 & 1.3 & 0.0145 & 11.41   & 314 & 747 & 5 \\ 
NGC~3621 & $7.2\pm0.3$ & 0.8 & 1.0 & 0.0079 & 10.77 & 186  & 2360 & 5 \\ \hline
NGC~6744 & $9.4\pm0.4$ & 0.7 & 2.4 &  0.0030 & 7.53   & 394 & 1435 & 6 \\
NGC~2775 & $23.2\pm3.5$ & 0.7 &  0.9  & 0.0043 & 7.98 & 275 & 198 & 6 \\
NGC~1433 & $18.6\pm1.9$  & 0.6 &  1.1  &	0.0029 & 10.77 & 192 & 357 & 6 \\
NGC~1512  & $18.8\pm1.9$  & 0.6 & 1.3 & 	0.0033 & 4.19 & 307 & 548 & 6 \\
NGC~2835 & $12.2\pm0.9$   & 0.6  & 1.3  & 0.0065 & 4.74 & 563 & 474 & 6 \\ \hline 
NGC~4298 & $14.9\pm1.4$ & 0.4 & 0.5 & 0.0057 & 16.45   & 273 & 466 & 7 \\
NGC~4689 & $15.0\pm2.3$ & 0.4 & 0.4 & 0.0040 & 9.18 & 341 & 308 & 7 \\
NGC~1317 & $19.1\pm0.8$ & 0.4 & 0.5 & 0.01945 & 38.37 & 34 & 49 & ...\\
NGC~685 & $19.9\pm3.0$ & 0.4 & 0.4 & 0.0025 & 3.61   & 299 & 227 & 7\\
IC~1954 & $12.8\pm2.1$ & 0.3 & 0.4 & 0.0091 & 12.16   & 154 & 206 & 7\\
NGC~4548 & $16.2\pm0.4$ & 0.3 & 0.5 & 0.0016 & 4.86   & 187 & 199 & 7\\ \hline
NGC~4571 & $14.9\pm1.2$ & 0.2 & 0.3 & 0.0030 & 7.53  & 159 & 142 & 8 \\
NGC~5068  & $5.2\pm0.2$   & 0.2  & 0.3 & 0.0059 & 5.76  & 182 & 630 & 8 \\ 
NGC~4826 & $4.4\pm0.2$ & 0.2 & 0.2 & 0.0068 & 11.99 & 172 & 121 & 8 \\
IC~5332  & $9.0\pm 0.4$ & 0.1  & 0.4 & 0.0015 & 0.95  & 229 & 182 & 8 \\  
\enddata
\tablecomments{Column~(2): distance in Mpc. Columns (3)–(4): Star formation rate in the HST footprint (SFR) and for the entire galaxy (SFR$_{\rm tot}$). The table is arranged from highest-to-lowest SFR. Column~(5): SFR per unit area.  Column~(6): surface density of molecular gas.  Columns~(7)-(8): Number of clusters in the primary, human-classified and secondary machine-learning classified catalogs. Column~(9): designated star-forming group number.}
\end{deluxetable*}

\begin{figure}[!htbp]
	\centering
	\includegraphics[width=\textwidth]{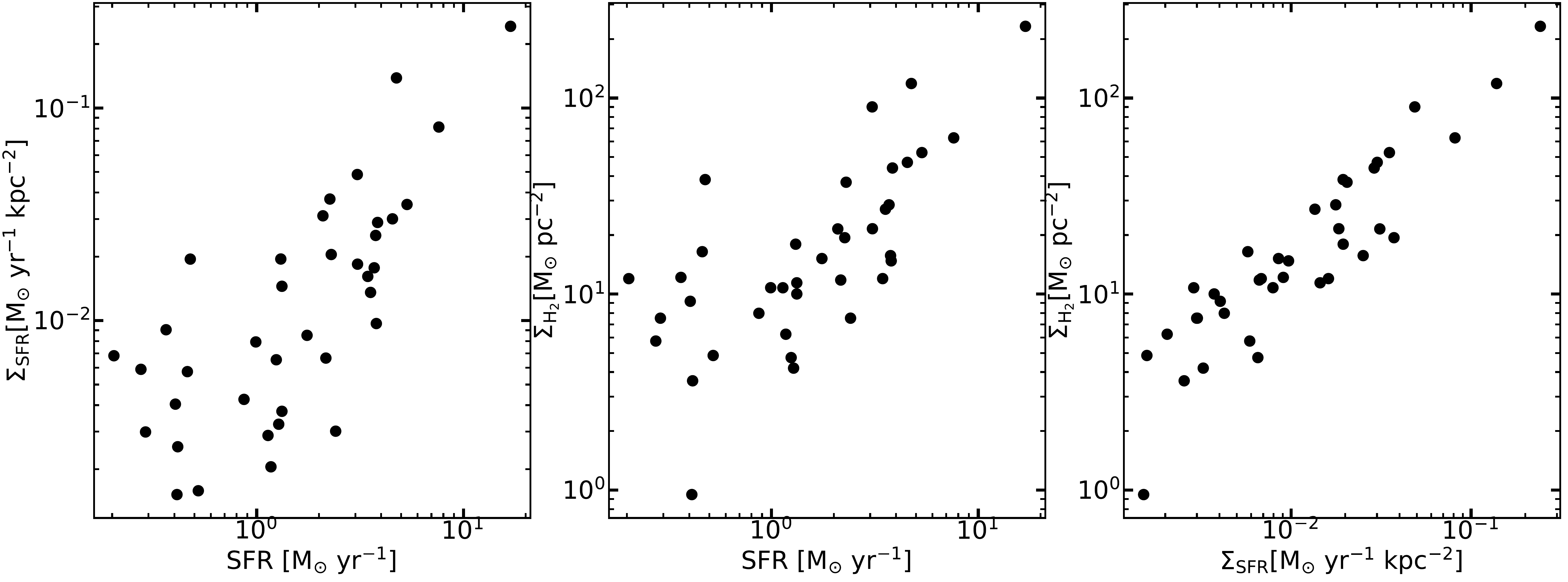}
\caption{{\bf Star-forming properties of PHANGS-HST spiral galaxies are well-correlated.} The star formation rate (SFR) in the HST footprint correlates well with the SFR per unit area $\Sigma_{\rm SFR}$ (left) and with the surface density of molecular gas $\Sigma_{\rm H_2}$ (middle) measured for the entire galaxy. $\Sigma_{\rm SFR}$ and $\Sigma_{\rm H_2}$ are also strongly correlated for our sample (right panel). \label{fig-sample}}
\end{figure}

Figure~\ref{fig-sample} demonstrates that the SFR in the HST footprint correlates well with $\Sigma_{\rm SFR}$ and $\Sigma_{\rm H_2}$ across the entire star-forming disk of the galaxy.
SFR and $\Sigma_{\rm SFR}$ (left panel) are well correlated, and both properties also correlate well with the surface density of molecular gas $\Sigma_{\rm H_2}$ (middle- and right- panels).

All galaxies in our sample were imaged in five broad-band filters (NUV, U, B, V, I).  Observation strategy is described in Section 3 of \citet{Lee2022} and exposure times for each target and filter are list in Table 1 \citet{maschmann24}. Ground-based H$\alpha$ imaging (A. Razza et al., submitted) and ALMA CO (2-1) maps \citep{phangs-alma} are also available for all galaxies.
The half of the sample with MUSE spectral maps (19 galaxies) was imaged at high-resolution in H$\alpha$ with HST as part of Cycle 30 Treasury program 17126 \citep{Chandar2025}, and in four near-infrared NIRCAM filters (F150W, F200W, F300M, and F335M) with JWST as part of Cycle~1 Treasury program 2107 \citep{phangs-jwst}.

One of the main goals of this work is to determine empirical mass and age distributions for cluster populations in spiral galaxies and to establish any relationship with star-forming properties of the host. To improve statistics, we create eight composite groups by combining together the cluster populations in 3 to 5 galaxies with similar SFRs. 
Groups 2-8 include galaxies with very similar SFRs. While the galaxies in Group~1 have a larger range in SFR, we found the results are insensitive to the  exact galaxies that are included, so elected to keep them together in one group. Results for the mass and age distributions of each galaxy individually are presented in Sections~4 and 5.
We also create three composite groups of five galaxies each with the highest, lowest, and intermediate $\Sigma_{\rm SFR}$ as an independent check.
Even for the composite groups the number of clusters is small in some of the age-mass bins. The composite cluster population in each group almost certainly represents near-constant formation over the past $\approx$Gyr, since variations in the rate of star and cluster formation are not expected to occur at similar times in different galaxies. The groups are arranged by decreasing star formation rate, identified in Table~\ref{tab:galaxysample}, and listed below.  

\begin{itemize}
\item {\bf Group~1 (SFR: $3.8-12.9~M_{\odot}~\mbox{yr}^{-1}$, total of $31.4~M_{\odot}~\mbox{yr}^{-1}$):}  NGC~1365, NGC~1672, NGC~4303, NGC~1097, and NGC~1559.  
\item {\bf Group~2 (SFR: $3.1-3.3~M_{\odot}~\mbox{yr}^{-1}$, total of $15.9~M_{\odot}~\mbox{yr}^{-1}$):} NGC~3627, NGC~1792, NGC~1566, NGC~4536, NGC~4654.
\item {\bf Group~3 (SFR: $2.0-2.4~M_{\odot}~\mbox{yr}^{-1}$, total of $8.8~M_{\odot}~\mbox{yr}^{-1}$):} NGC~4321, NGC~2903, NGC~1385, NGC~7496. 
\item {\bf Group~4 (SFR: $1.0-1.3~M_{\odot}~\mbox{yr}^{-1}$, total of $4.7~M_{\odot}~\mbox{yr}^{-1}$):} NGC~1087, NGC~4569, NGC~4535, NGC~1300. 
 \item {\bf Group~5 (SFR: $0.8-0.9~M_{\odot}~\mbox{yr}^{-1}$, total of $2.6~M_{\odot}~\mbox{yr}^{-1}$):} NGC~628, NGC~3351, NGC~3621.  
\item {\bf Group~6 (SFR: $0.6-0.7~M_{\odot}~\mbox{yr}^{-1}$, total of $3.2~M_{\odot}~\mbox{yr}^{-1}$):} NGC~6744, NGC~2775, NGC~1433, NGC~1512, NGC~2835.  
 \item {\bf Group~7 (SFR: $0.3-0.4~M_{\odot}~\mbox{yr}^{-1}$, total of $1.8~M_{\odot}~\mbox{yr}^{-1}$):} NGC~4298, NGC~4689, NGC~685, IC~1954, NGC~4548. 
 \item {\bf Group~8 (SFR: $0.1-0.2~M_{\odot}~\mbox{yr}^{-1}$, total of $0.7~M_{\odot}~\mbox{yr}^{-1}$):} NGC~4571, NGC~5068, NGC~4826, IC~5332.  
\end{itemize}
The cluster populations of two galaxies are not included in any of the groups (but are studied as individual galaxies) because their star formation rates do not fall neatly into any of the ranges: NGC~4254 ($2.8~M_{\odot}~\mbox{yr}^{-1}$) and NGC~5248 ($1.7~M_{\odot}~\mbox{yr}^{-1}$). NGC~1317 is not included in a group because it has a very small cluster sample.  

\subsection{Star Cluster Catalogs}

In this work, we use age and mass estimates for compact stellar clusters in 38 spiral galaxies from the catalogs presented in \citet{Thilker25} and \citet{Henny25}.

\subsubsection{Cluster Selection and Improvements to Age-Dating}
The PHANGS team has released new, state-of-the-art catalogs of compact star clusters for 38 spiral galaxies\footnote{\url{https://archive.stsci.edu/hlsp/phangs/phangs-cat}}. These catalogs are the product of a series of efforts, as summarized in \citet{Lee2022}, to improve techniques for cluster selection \citep{whitmore21, Thilker22}, aperture corrections \citep{deger22}, and automated morphological classification using machine-learning (ML) techniques \citep{wei20, whitmore21, hannon23}.  We briefly summarize the selection process here; full details can be found in \citet{Thilker22}.
For each galaxy, an initial list of extended sources are automatically selected  from their structural parameters, followed by classification based on their morphology in the V band. Class `1' and `2' sources have a single, centrally concentrated peak that appears symmetric or slightly asymmetric; these are the clusters we use in this work. For very young ($< 10$~Myr) clusters in particular, individual bright, massive and/or evolved stars can result in an asymmetric appearance regardless of the spatial distribution of the more numerous, lower-mass stars.
Sources with multiple peaks are designated class `3' compact associations and excluded from this work. These morphological classifications do not tell us unambiguously if a cluster is gravitationally bound (has negative energy) or unbound (positive energy), although the bound fraction may depend at least somewhat on the class \citep[e.g.,][]{Kruijssen12,Ginsburg16,Grudic21}. While sources older than $\approx10$~Myr are almost certainly gravitationally bound \citep[e.g.,][]{Gieles11}, candidates younger than this contain an unknown mix of bound clusters and unbound associations. We do not attempt to distinguish bound from unbound sources based on morphology,
rather we will report fitting results with and without clusters younger than 10~Myr.

The PHANGS-HST cluster catalogs includes the results of two separate source classification efforts, where source morphology was assessed by a human \citep[][N$\sim$15,000]{whitmore21} and another by a machine-learning algorithm \citep[][N$\sim$30,000]{wei20,hannon23}, and we present results from both here. Sample sizes are reproduced in Table~1 in this paper from \citet[][Table 2]{Maschmann2024}.
The {\bf primary} cluster catalogs used in this work are those identified as class $1+2$ by a human. These catalogs have been vetted in a number of works and contain minimal contamination. However, they are time-intensive to produce, resulting in a brighter completeness limit than the machine-learning based catalogs for most galaxies \citep[][Figure 3]{Maschmann2024}. 
We also use the machine-learning based class $1+2$ sources as secondary catalogs. These catalogs were produced using the VGG19 neural network architecture (details in \citealt{wei20}), and have the advantages of being reproducible (given the same model) and larger, at a cost of somewhat more contaminants. 
\citet{whitmore21} and \citet{hannon23} made detailed comparisons between the human- and machine learning-classified cluster catalogs.  They found that  Class~1+2 sources (those used in this work) agree at the 80-85\% level between the two classification methods,  making the neural network classifications comparable in quality to the human-based cluster catalogs.

Two well-known challenges when estimating the age and mass of star clusters from 
their spectral energy distributions (SEDs) are: (1) for clusters younger than a few$\times100$~Myr, the effects of age and reddening are degenerate in broadband colors \citep[e.g.,][]{anders04a,lsmith07,chandar10c}, and (2) for clusters older than $\approx1$~Gyr broadband colors are primarily sensitive to metallicity rather than to age \citep[e.g.,][]{Worthey94}. The PHANGS team has recently made important improvements to their age-dating procedures to address these issues. For the catalogs used here the age, reddening, and mass of each cluster were estimated by fitting its observed SED with predictions from the single-age solar metallicity cluster evolution models of \cite{Bruzual2003} within the CIGALE fitting framework \citep{Boquien2019}. The models range in age from 1~Myr through 12~Gyr, with linear sampling between 1 and 10~Myr and even sampling in log space after, and the fitting allows for a V-band extinction A$_V$ up to 5~mag with an assumed Milky-Way-like extinction curve \citep{fitzpatrick99}.

The primary catalogs from \citet{Thilker25} use a consistent method to estimate the ages and masses of clusters across all 38 galaxies. In addition to the measured cluster SEDs, they use information about H$\alpha$ emission, source location relative to spiral arms, and observed U$-$B vs V$-$I colors to determine whether an object is very young with nebular emission or an old globular cluster, then fit these two groups with different model grids. Detailed checks show that the erroneously young age estimates for old globular clusters produced by the original pipeline \citep[][]{Whitmore23,Floyd24} are fixed in the new catalogs, and that establishing whether or not a cluster has H$\alpha$ emission significantly improves the age estimates for young clusters. Details of the age-dating method and a number of checks can be found in \citet{Thilker25}.

Secondary catalogs with additional improvements to both the measured SEDs and model predictions are available for a subset of 16 galaxies \citep{Henny25}. This work makes use of photometry in 10 filters to construct the SED of each cluster by adding H$\alpha$ (from HST) and four-filter near-infrared photometry (from JWST) to the five broad-band measurements from HST. Adding these filters directly to the SEDs helps to determine which of the two or three combinations of age and reddening are correct for each cluster. Details of the fitting procedure, which include improvements to the \cite{Bruzual2003} models to account for gas and dust emission, are presented in \citet{Henny25}. Unfortunately these improved cluster ages and masses are currently only available for a subset of clusters in the 16 galaxies, since the procedure requires photometry in all 10 filters, and the HST-H$\alpha$ and JWST/NIRCAM footprints do not perfectly match the broad-band HST filters.

 Undoubtedly some biases remain despite these improvements to cluster age dating in the primary and secondary catalogs. In particular, 
distinguishing $\approx7-10$~Myr clusters from those with ages between $\approx20-50$~Myr is challenging from integrated photometry alone.  In the future it would be worthwhile to explore whether integrated spectra provide clues for additional improvements to the age-dating methodology. See \citet{Krumholz19} for additional discussion of this point.

\subsubsection{An Assessment of Incompleteness}

Incompleteness can artificially flatten the low end of the cluster mass function.
Correcting cluster catalogs in spiral galaxies for incompleteness is quite challenging, since the completeness percentage for every cluster is a complicated function of brightness, size, crowding, local background level, and distance to the galaxy. We use a conservative approach and restrict our analysis to masses where incompleteness has little impact on the observed distributions. Observed cluster mass functions follow a power law but eventually flatten towards lower masses due to incompleteness. Following the technique developed by \citet{Mok20}, we take the lower mass limit M$_{\rm lim}$ in a given age interval to be the mass where the cumulative distribution falls below the extrapolated power law by a factor of two. This approach should work well for the cluster populations in most galaxies in our sample.

The largest uncertainties are for the youngest ($\lea 10$~Myr) cluster populations in the galaxies with the highest SFRs, particularly NGC~1365 and NGC~1672, which are forming many clusters in crowded, dusty regions. In these two galaxies, we restrict the mass function fits for very young $<10$~Myr clusters to stay above $10^5~M_{\odot}$. For NGC~1672 the cluster mass function has a power-law index of $-1.95$, which is quite close to $-2$, the index found for clusters in the disks of Local Group galaxies (see \citealt{Krumholz19} and references therein).
This suggests that incompleteness does not significantly affect the high-mass end. For NGC~1365 the cluster mass function is somewhat flatter, with a power-law index of $-1.56$ even when the fit is restricted to clusters more massive than $10^5~M_{\odot}$. It is possible that the initial cluster mass function in this galaxy is shallower than the canonical power-law index of $-2$, as predicted for regions of intense star formation \citep{Li22}, or that incompleteness affects the sample even at these high masses.

\section{Cluster Age-Mass Distributions}\label{sec-Mt}

\begin{figure}[!htbp]
	\centering
\includegraphics[width=18cm]{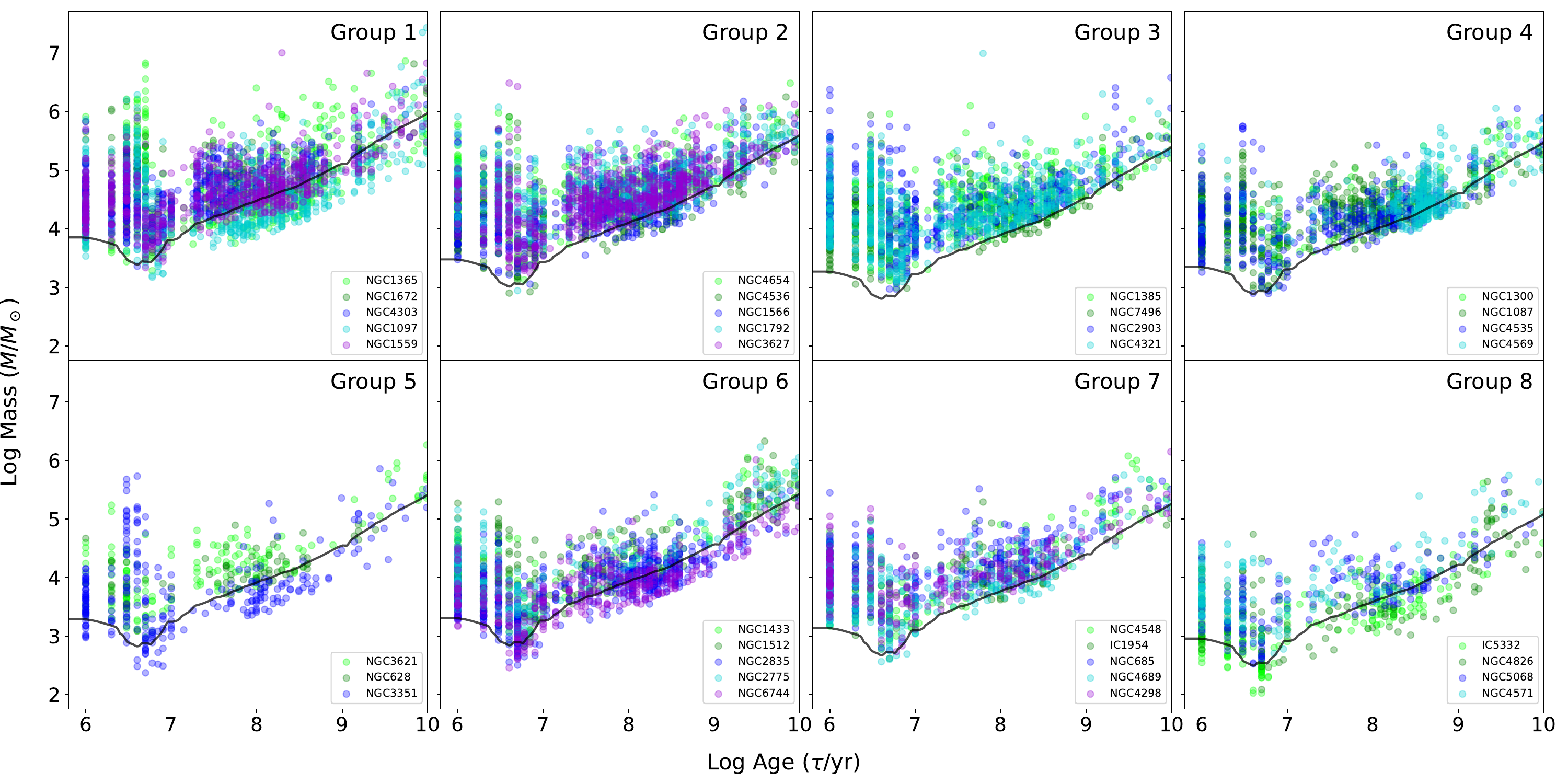}
	\caption{{\bf The age-mass diagrams of clusters in the eight galaxy groups} that are used throughout this work and described in Section~2.1, organized from highest to lowest SFR. The cluster populations formed continuously over at least the past Gyr, and the composite SFR was likely approximately constant over this period since multiple galaxies have been stacked together and it is unlikely they experienced variations in their SFR at the exact same time. The decrease in the upper end of cluster masses with SFR is quite noticeable, with a drop by a factor of $\approx100$ from the highest to lowest star-forming group. The black line represents the luminosity limit in the galaxy with the most conservative lower mass limit within the group.}\label{fig-Mt_SFgroups}
\end{figure}

The mass and age distributions of cluster populations in galaxies provide direct insight into their formation and evolution. Figure~\ref{fig-Mt_SFgroups} plots log$M$ vs. log$\tau$ for the cluster populations in the eight galaxy groups, starting from the group with the highest star formation rates in the top-left panel to the lowest in the bottom-right panel.

Clusters in all panels span the full range of ages from $\approx$~1 Myr to $\approx10$~Gyr. The mass ranges however, particularly at the upper end, are different, with clusters in the top-left panel (highest SFR) reaching masses approaching $10^7~M_{\odot}$ while those in the bottom panel (lowest SFR) only approach $10^5~M_{\odot}$. In fact, the most massive clusters in our sample appear to scale directly with the SFR of the host galaxy (see Section~6.4). We will quantify the observed relationship between the cluster mass function and host galaxy SFR in an upcoming work.

While young, non-nuclear star clusters significantly more massive than $\approx10^5~M_{\odot}$ are expected to be rare in spiral galaxies \citep[e.g.,][]{portegieszwart2010}, our mass-age distributions {\em suggest that clusters with masses of $\approx10^6~M_{\odot}$ and higher can form in spiral galaxies with high rates of star formation.} This is not an artifact or bias from the age-dating procedure. Several works have studied the properties of very massive, embedded clusters in NGC~1365 \citep[][]{Kristen97,Galliano08,Whitmore23b}, and we have inspected each cluster with $M > 10^5~M_{\odot}$ and log~$(\tau/\mbox{yr})<6.6$ to confirm that it is in fact young and massive.

The prominent lower edge in each panel is the luminosity limit for the cluster population in each galaxy. Because clusters fade significantly (by $\approx5$~mag) over the first 10~Myr as the brightest, most massive stars die out, luminosity-limited samples do not include lower-mass clusters at older ages. The solid black line in each panel shows the most conservative luminosity limit for each group (i.e. for the galaxy with the brightest M$_V$ limit to its cluster population). Cluster mass and age distributions presented in the following sections for the galaxy groups are restricted to stay above this limit.

In Figure~\ref{fig-Mt_SFgroups} the data points `pile-up' at certain ages and show small `gaps' at others. This occurs at ages where the uncertainties in log~$\tau$ are not symmetric. For example there is a prominent gap in the range log~$(\tau/\mbox{yr}) \approx 7.0-7.3$, where the optical emission from massive clusters is dominated by red supergiant stars and the predicted cluster colors loop back on themselves. Here, the fitted ages become degenerate and tend to avoid values near log~$\tau \approx 7.0$. There may be weak features like this in the unbinned age distribution that are not present in the real age distribution as a result of this type of bias, and the locations of such features likely differ when different age-dating methods are used. Overall, the broad distribution of cluster ages (and masses) are not greatly affected by these types of small-scale features.

Much of the statistical information about cluster populations can be seen, at least qualitatively, from the log$M$ vs. log$\tau$ plots. In each panel, when we look down the vertical axis we see that the number of clusters increases with decreasing mass. This suggests that if the clusters follow a power law-like mass distribution, $dN/dM \propto M^\beta$, the index must be steeper than $-1$, i.e. $\beta < -1$. Clusters in galaxies with higher SFRs tend to have higher masses.  The fact that the most massive clusters scale with SFR (see more in Section~6.4) indicates cluster masses are consistent with being randomly sampled from a power-law, where galaxies with higher SFRs form more clusters and therefore sample the power-law more frequently resulting in more massive clusters than in galaxies with lower SFRs.  
When we look horizontally along a given mass interval, the number of clusters is approximately equal or increases only gradually in equal bins of log~$(\tau/\mbox{yr})$, without piling up or dropping off sharply at older ages. This means that if the clusters have a power law-like age distribution, $dN/d\tau \propto \tau^{\gamma}$, then $\gamma \approx -1$. We quantify mass distributions in Section~4 and age distributions in Section~5.

\section{Cluster Mass Functions}\label{sec-cmf}

\subsection{Methodology}

One of the simplest and most informative approaches for investigating the evolution of stellar clusters is from averages of their mass-age distributions $g(M,\tau)$ over several adjacent intervals of age and mass. This approach makes it easy to identify changes to the shape or normalization of the distribution. The mass function for a given age interval can be calculated as:
\begin{equation}
    \overline{g}(M) \equiv \frac{1}{(\tau_2-\tau_1)}\int_{\tau_1}^{\tau_2} g(M,\tau)d\tau,
\end{equation}
where $\tau_1$ and $\tau_2$ represent the youngest and oldest age included in the interval.
For a population of clusters that forms at a constant rate and experiences {\em no} dissolution, the mass function will have the same shape and amplitude in different age intervals (i.e. the $\overline{g}(M)$ distributions will fall on top of one another).  If mass-dependent dissolution dominates the evolution, the high mass end of the distributions will still lie on top of one another but older distributions will flatten at lower masses. 
And if mass-{\em in}dependent dissolution dominates cluster evolution, the shapes of the distributions will not change, but their amplitudes will decrease over time.

We will present cluster mass functions in four successive age intervals:
\begin{itemize}
\item  $6.0 \leq$ log~$(\tau/\mbox{yr}) < 7.0$ ($1-10$~Myr)
\item  $7.0 \leq$ log~$(\tau/\mbox{yr}) < 8.0$ ($10-100$~Myr)
\item  $8.0 \leq$ log~$(\tau/\mbox{yr}) < 8.6$ ($100-400$~Myr)
\item  $8.6 \leq$ log~$(\tau/\mbox{yr}) \leq 9.0$ ($400 - 1000$~Myr)
\end{itemize}
The first three intervals have been used in a number of previous works and reasonably account for common age-dating biases. For example binning together clusters with estimated ages between 10 and 100~Myr minimizes the impact of the `gap' discussed in Section~3.
The oldest interval has not been used extensively in studies of extragalactic cluster populations. We include it here because the improved treatment of globular clusters by the PHANGS collaboration prevents these ancient stellar systems from being incorrectly dated to ages in this range \citep[][]{Thilker25,Henny25}.
We do not study clusters older than 1~Gyr (log~$(\tau/\mbox{yr}) > 9.0$) because it is not possible to establish their ages with any certainty based on integrated colors alone, 
and because at least some fraction may be subject to different physical processes because they reside in the halo rather than the disk.


\begin{figure}[!htbp]
	\centering
	\includegraphics[width=\linewidth]{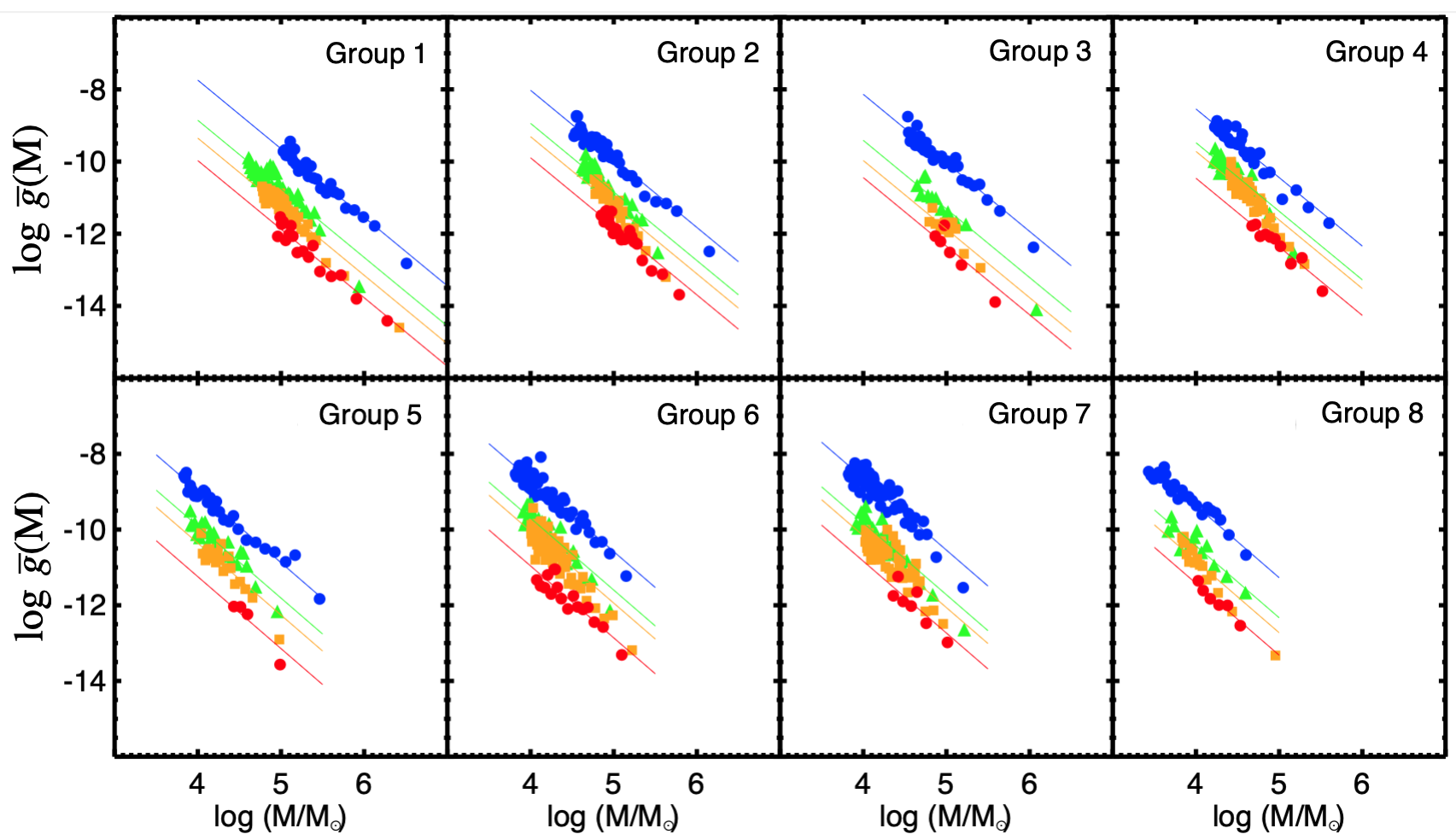}
	\caption{{\bf Mass function of star clusters for the eight galaxy groups have similar shapes but different amplitudes.} The groups are arranged in order of decreasing star formation rate starting from the upper-left.  Each panel shows the mass function in four intervals of age: $1-10$~Myr (blue), $10-100$~Myr (green), $100-400$~Myr (orange), and $0.4-1$~Gyr (red).  Each distribution is divided by the amount of time elapsed in the interval, and the normalizations are preserved.   The lines show that a power-law with index $\beta=-1.9$ is a good approximation to each distribution; the best fit values of $\beta$ are compiled in Table~\ref{tab:betagroups}. The normalization of the mass functions decreases systematically with increasing cluster age in all galaxy groups. 
    }.\label{fig-gM_groups}
\end{figure}

In Figure~\ref{fig-gM_groups} we present the mass functions $\overline{g}(M)$ based on our primary cluster catalogs in the eight galaxy groups, arranged from  highest to lowest star formation rate starting from the top-left.
Distributions for 1-10~Myr clusters are plotted in blue, 10-100~Myr in green, 100-400~Myr in orange, and 400~Myr-1~Gyr in red. The distributions cover different mass ranges, with the highest star-forming galaxies (group~1, top-left) extending to masses above $10^6~M_{\odot}$ but also having among the highest mass limits ($\approx3\times10^4~M_{\odot}$), while the lowest star-forming galaxies (group 8, low-right) barely reach $3\times10^4~M_{\odot}$ but extend down to $\approx3\times10^3~M_{\odot}$. Together, the distributions cover a factor $\approx10^3$ in mass from $\sim 3\times10^3~M_{\odot}$ to $\sim 3\times10^6~M_{\odot}$.

We construct the mass function of clusters in the four age intervals listed above using variable-size bins with equal numbers of clusters in each bin; this allows reliable fits to the distributions with little bias in the power-law index, even in cases with few clusters \citep[e.g.,][]{maiz05}. 
The completeness limit M$_{\rm lim}$ for each distribution is determined quantitatively from its cumulative mass distribution. As explained in Section~2.2.2,  cluster mass functions follow a power law at the high-mass end, but eventually flatten toward lower masses due to incompleteness. Following the method developed by \citet{Mok20}, we define the lower mass limit M$_{\rm lim}$ for each distribution to be where the distribution falls below the extrapolated power law by a factor of two.

The mass functions plotted in Figure~\ref{fig-gM_groups} all appear fairly linear, suggesting they can be approximated by a single power law.
The distributions have roughly the same slope in the different age intervals and across the galaxy groups, with each distribution increasing continuously towards lower masses. 
We determine the best-fit power-law index $\beta$ for each distribution from a least-squares fit of the form log~$\overline{g}(M) = \beta \mbox{log}M +$const over the plotted mass range, along with $1\sigma$ uncertainties, and record the results in Table~\ref{tab:betagroups}.\footnote{Testing whether or not the cluster mass functions deviate downward from a power-law at the high-mass end (as has been suggested by some studies) is beyond the scope of this work and will be explored in a future paper.}
The cluster mass functions for individual galaxies are fitted using the same procedure, and the results are presented in Table~\ref{tab:betagals} and Figures~\ref{fig-gM_gals1} and \ref{fig-gM_gals2} in the Appendix.


\begin{deluxetable*}{lccccccccc}
\tablecaption{Power-law Index $\beta$ and offsets for Mass Functions in Composite Galaxy Groups\label{tab:betagroups}}
\centering
\tablehead{
\colhead{Group} & SFR & \colhead{$\beta1$} & \colhead{$\beta$2} & \colhead{$\beta3$} &  \colhead{$\beta4$}  & \colhead{$\Delta \beta_{41}$\tablenotemark{a}} & \colhead{$\Delta \beta_{42}$\tablenotemark{a}} & \colhead{$\Delta gM_{41}$\tablenotemark{b}} & \colhead{$\Delta gM_{42}$\tablenotemark{b}} \\ 
 \colhead{\#} & ($M_{\odot}~\mbox{yr}^{-1}$) & \colhead{ (1-10~Myr)}  &
 \colhead{(10-100~Myr) } & \colhead{(100-400~Myr) } & \colhead{(0.4-1~Gyr) }  & \colhead{} & \colhead{ } & \colhead{ } }
\startdata
1 & 3.7-12.7 & $-2.02\pm0.08$ & $-2.33\pm0.13$ & $-2.27\pm0.09$ & $-2.06\pm0.15$ & $-0.04\pm0.17$ & $0.27\pm0.20 $ & 2.14 & 1.13   \\ 
2 & 3.0-3.3 & $-1.92\pm0.07$ & $-2.19\pm0.09$ & $-2.09\pm0.31*$ & $-2.45\pm0.25$ & $-0.53\pm0.26$ & $-0.26\pm0.26$ &  1.87 & 0.95 \\ 
3  & 2.0-2.4 & $-1.88\pm0.08$ & $-2.39\pm0.22$ & $-2.35\pm0.45$ & $-2.55\pm0.43$ & $-0.67\pm0.44$ & $-0.16\pm0.47$ & 2.31 & 1.03 \\
4 & 1.0-1.25 & $-1.90\pm0.09$ & $-2.02\pm0.14$ & $-2.14\pm0.11$ & $-2.08\pm0.13$  & $-0.18\pm0.16$  & $-0.06\pm0.21$ & 1.92 & 0.99 \\
5 & 0.8-0.9 & $-1.84\pm0.08$ & $-2.21\pm0.22$ & $-2.45\pm0.24$ & ... & ... & ... & 2.27 & 1.33 \\
6 & 0.6-0.7 & $-2.05\pm0.11$ & $-2.43\pm0.12$ & $-2.36\pm0.24$ & $-2.15\pm0.28$ & $-0.10\pm0.30$ & $0.28\pm0.30$ & 2.28 & 1.26 \\
7 & 0.3-0.4 & $-2.20\pm0.15$ & $-2.32\pm0.12$ & $-2.17\pm0.24$ & $-2.08\pm0.38$ & $+0.12\pm0.41$ & $0.24\pm0.40 $ & 2.19 & 1.01 \\
8 & 0.1-0.2 & $-2.04\pm0.36$ & $-2.55\pm0.24$ & $-2.11\pm0.12$ & $-2.05\pm0.26$ & $-0.01\pm0.44$ & $0.50\pm0.35$ & 2.05 & 0.99 \\
\enddata
\tablenotetext{a}{The difference in $\beta$ values measured for clusters in the $0.4-1$~Gyr and $1-10$~Myr intervals ($\Delta \beta_{41}$) and the $0.4-1$~Gyr and $10-100$~Myr intervals ($\Delta \beta_{42}$). Positive values indicate a cluster mass function that flattens over time; none of the results are consistent with this scenario. }
\tablenotetext{b}{The difference in amplitude between the cluster mass function measured for clusters in the $0.4-1$~Gyr and $1-10$~Myr ($\Delta gM_{41}$) and $0.4-1$~Gyr and $10-100$~Myr intervals ($\Delta gM_{42}$). The uncertainties are approximately $\pm 0.1$ (see text).}
\end{deluxetable*}

Besides the striking similarity of the slopes, one of the most noticeable features in Figure~\ref{fig-gM_groups} is the continuous decrease in amplitude with increasing age, i.e. the downward trend from blue to green to orange to red. This systematic decrease is observed in all eight groups. We quantify this drop, which provides constraints on the rate of cluster dissolution over time, by fitting a power-law with a fixed slope of $\beta=-1.9$ (see below) to each cluster mass function, and calculating the y-intercept value. We compile the difference in y-intercept values between the youngest ($1-10$~Myr) and oldest (400~Myr$-$1~Gyr) distributions, $\Delta g_{41}$ in column~8 of Table~\ref{tab:betagroups}, and the difference between the second youngest ($10-100$~Myr) and oldest distributions $\Delta gM_{42}$ in column~9. These values 
are not very sensitive to the exact value of $\beta$ that we adopt, and do not vary by more than $\pm0.1$ if $\beta=-1.7$ or $-2.1$ are assumed instead. 

\subsection{Results}

The best-fit power-law index $\beta$ (columns~3-6 in Table~\ref{tab:betagroups}) ranges from $\approx-1.8$ to $\approx-2.5$ for the galaxy groups, although some of the fits have large formal uncertainties. 
For the `initial' cluster mass function, i.e. $1-10$~Myr, the median power-law index is $\beta =-1.92$, with a mean and standard deviation of $-1.9\pm0.1$. In the oldest $0.4-1$~Gyr interval, the median value is $\beta=-2.2$, with a mean and standard deviation of $-2.1\pm0.2$. These results suggest that {\em the power-law index of the cluster mass function is the same within the fit uncertainties} for the first $\approx1$~Gyr, at least over the plotted mass ranges.

We make a few checks of the results. The power-law index $\beta$ is also determined from a Maximum Likelihood fit, which makes no assumptions on binning. We find these fits for $\beta$ are within $1\sigma$ of the results in Table~\ref{tab:betagroups}.  Tests using different (higher) mass limits for the distributions and different numbers of clusters per bin also do not significantly impact $\beta$ except in cases with poor statistics. 

\begin{figure}[!htbp]
	\centering
\includegraphics[width=14cm]{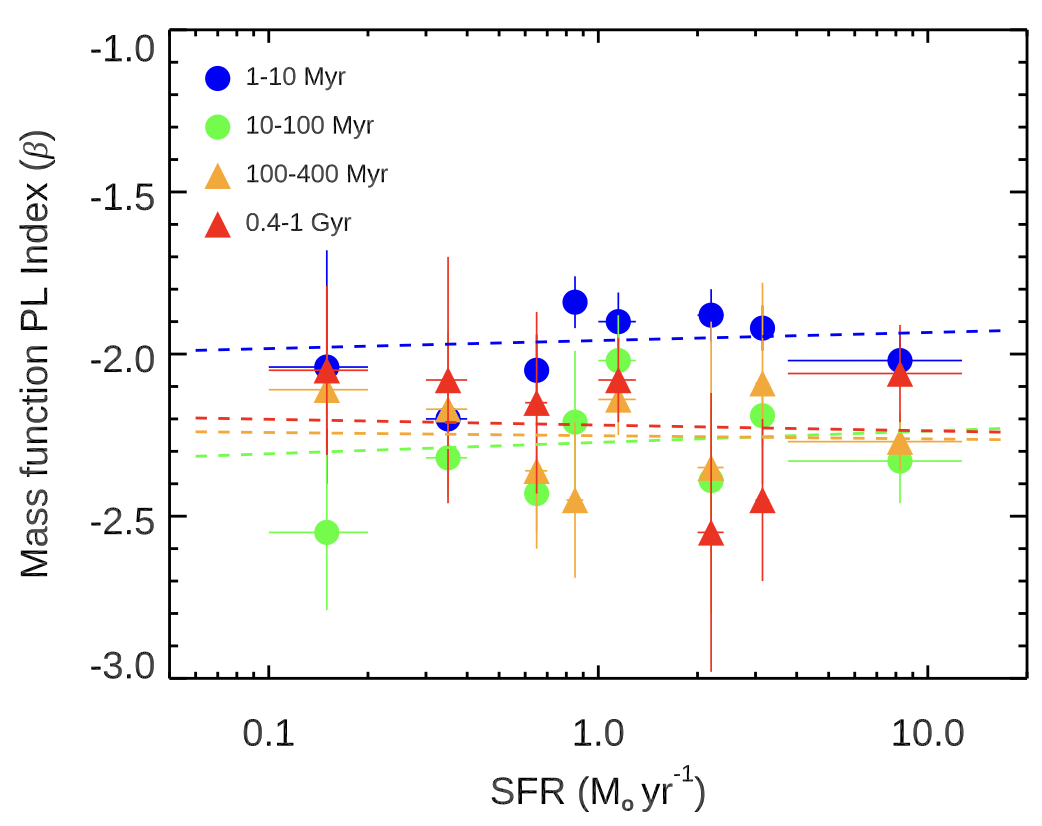} 
	\caption{{\bf There is no statistically significant trend between the power-law index $\beta$ for the cluster mass function and the SFR of the galaxy groups. } The best fit power-law index $\beta$ to the observed cluster mass functions for the eight galaxy groups are shown for the same four age intervals as Figure~\ref{fig-gM_groups} as indicated in the legend. There are no statistically significant trends between $\beta$ with age or SFR for the composite galaxy groups.   
    }\label{fig-betavssfr}
\end{figure}

In Figure~\ref{fig-betavssfr} we plot the results for $\beta$ color-coded by age interval as indicated in the legend, versus the mean star-formation rate of each galaxy group.
The dashed lines show that there is no statistically significant trend between $\beta$ and SFR for any age interval, with best linear fits of: $0.004\pm0.018$ (1-10~Myr), $0.005\pm0.025$ (10-100~Myr), $-0.001\pm0.020$ (100-400~Myr), and $-0.003\pm0.032$ (400~Myr-1~Gyr). While the youngest clusters have slightly shallower mass functions, their formal fits for $\beta$ are within $1\sigma$ of the best fit values for the other three age intervals. Results for $\beta$ determined from the machine-learning selected cluster catalogs for the exact same galaxy groups, age intervals, and mass limits also show no correlation between $\beta$ and SFR.

\begin{figure}[!htbp]
	\centering
\includegraphics[width=\textwidth]{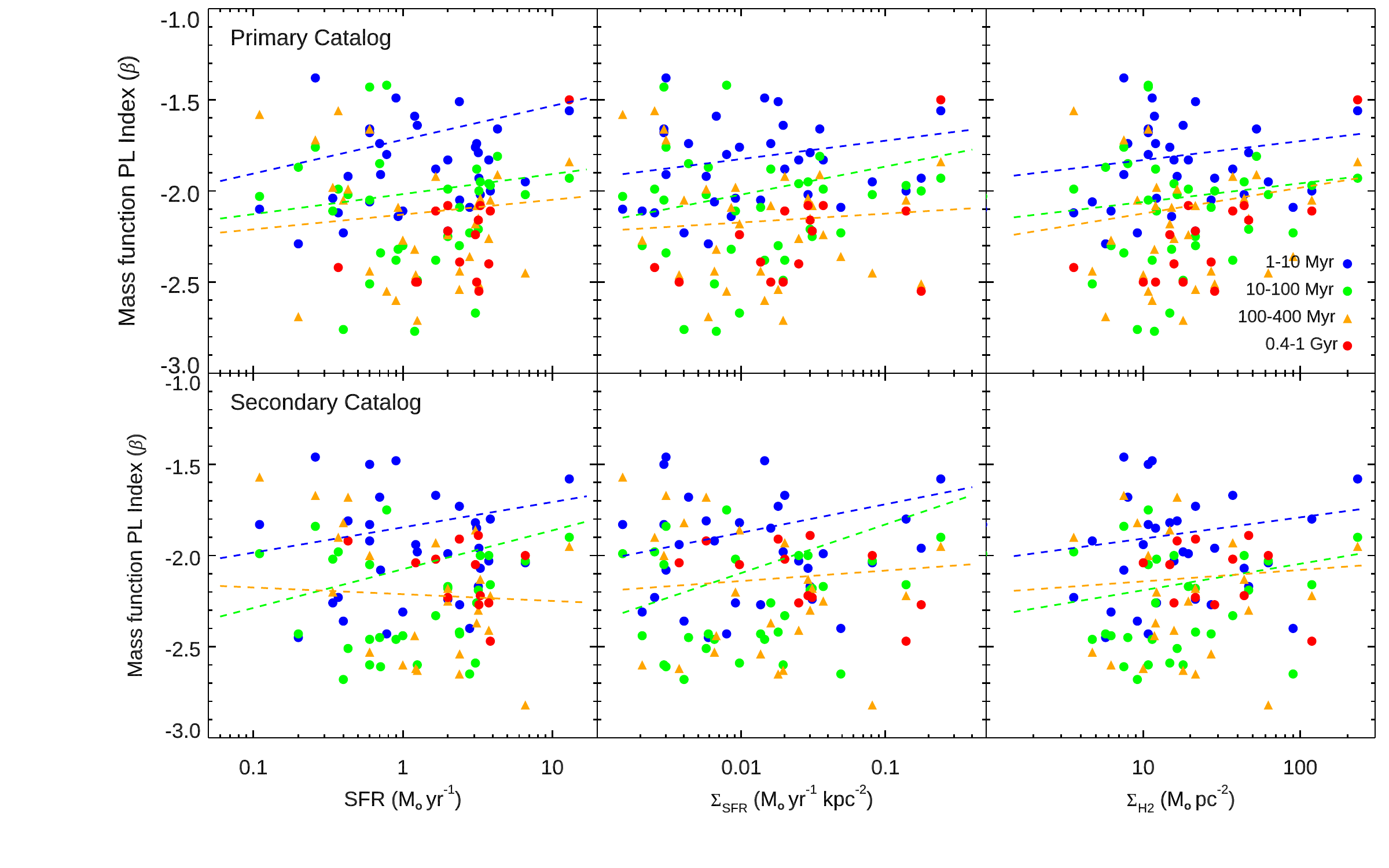}
	\caption{{\bf There are no statistically significant trends between the mass function power-law index, $\beta$, for cluster populations in individual galaxies and star-forming properties of the host. } The best fit power-law index $\beta$ to the observed cluster mass functions for all 38 individual galaxies are plotted against the SFR, $\Sigma_{\rm SFR}$, and $\Sigma_{H_2}$ for the host galaxy {\bf from the primary (human-classified) (top panels) and secondary (machine learning-classified) cluster catalogs (bottom panels).}  
    The age intervals are indicated in the legend. Several of the individual galaxies have poor statistics in the 0.4-1~Gyr interval. }\label{fig-betagals}
\end{figure}

The best-fit $\beta$ values for individual galaxies are plotted in Figure~\ref{fig-betagals} with the same color-coding for the age intervals as the previous figure,  including results from the primary (top panels) and secondary (bottom panels) cluster catalogs.
The results are similar to those for the galaxy groups {\bf and between the two catalogs}, but exhibit larger scatter and one possible exception. There is a weak trend for $\beta$ for very young $1-10$~Myr clusters to be shallower for galaxies with high SFR, represented by the dashed blue line in the left panel of Figure~\ref{fig-betagals}. The linear fit has a slope of $0.027 \pm 0.015$, a marginal trend significant only at the $\approx1.5\sigma$ level.
Simulations \citep{Li22} and some observations (see \citealt{Krumholz19} and references therein) suggest the mass function for clusters will be flatter when they form at high pressure, which is expected in galaxies with high rates of star formation. Note however that the power-law index for the mass functions of $10-100$~Myr, $100-400$~Myr, and $0.4-1$~Gyr clusters do not show a statistically significant trend with SFR (the slopes of the fits are $0.025\pm0.024$, $0.012\pm0.025$, and $0.003\pm0.043$, respectively). We find a similar lack of correlation for $\beta$ and composite galaxy groups with high, medium, and low $\Sigma_{\rm SFR}$. 
 {\em We adopt the mean and standard deviation of $\overline{\beta}=-1.9\pm0.2$} from the initial cluster mass function to characterize our sample.
 
We find no statistically significant trend between $\beta$ and $\Sigma_{\rm SFR}$ or $\Sigma_{\rm H_2}$ for any age interval in the individual galaxies, as shown in the middle and right panels of Figure~\ref{fig-betagals}.
The open symbols represent $\beta$ values determined for the galaxies in the \citet{Henny25} sample, and are qualitatively similar to the results from the primary catalogs (see Figure~\ref{fig-gMKiana}, Table~\ref{tab:betahenny} in the Appendix). Because these can have significantly fewer clusters with age and mass estimates than in the full catalogs, we mainly use the \citet{Henny25} catalogs as an independent check on our results for spiral galaxies. 

While the {\em shapes} of the cluster mass functions do not evolve over the first $\approx$Gyr, the {\em amplitudes} show a continuous decrease over time. We quantify this decrease by measurements $\Delta gM_{41}$ and $\Delta gM_{42}$, which we compile for the composite groups in Columns~9 and 10 of Table~\ref{tab:betagroups}. The mass functions for all groups show a similar decrease in amplitude over time, with a mean$+$ standard deviation of $2.13\pm0.17$ for $\Delta gM_{41}$ and $1.09\pm0.14$ for $\Delta gM_{42}$. There is no trend in either statistic with SFR. The cluster mass functions of individual galaxies show a similar continuous decrease in amplitude (Appendix).

\section{Age Distributions}\label{sec-dndt}

\subsection{Methodology}

Age distributions are created by counting the number of clusters within a given mass interval and age bin, and dividing by the amount of time covered by the bin. 
The $1\sigma$ uncertainties are determined as $\pm \sqrt{N}$.
The age distributions are then normalized by the mass interval:
\begin{equation}
    \overline{g}(\tau) \equiv \frac{1}{(M_2-M_1)}\int_{M_1}^{M_2} g(M,\tau)dM,
\end{equation}
where M$_2$ and M$_1$ are the highest and lowest masses used to create the age distribution.

We determine the age distribution in three adjacent mass intervals for the cluster populations in each galaxy group to assess whether there are systematic changes in its shape. The specific mass intervals vary from group-to-group. We restrict one interval in each group to sufficiently high masses that the distribution remains above the luminosity limit in Figure~\ref{fig-Mt_SFgroups} to log~$(\tau/{\rm yr})=9.0$. The other two distributions are constructed from adjacent intervals of lower mass clusters that stay above the luminosity limit to an age of log~$(\tau/{\rm yr})=8.4$.

\begin{figure}[!htbp]
	\centering
	\includegraphics[width=18.0cm]{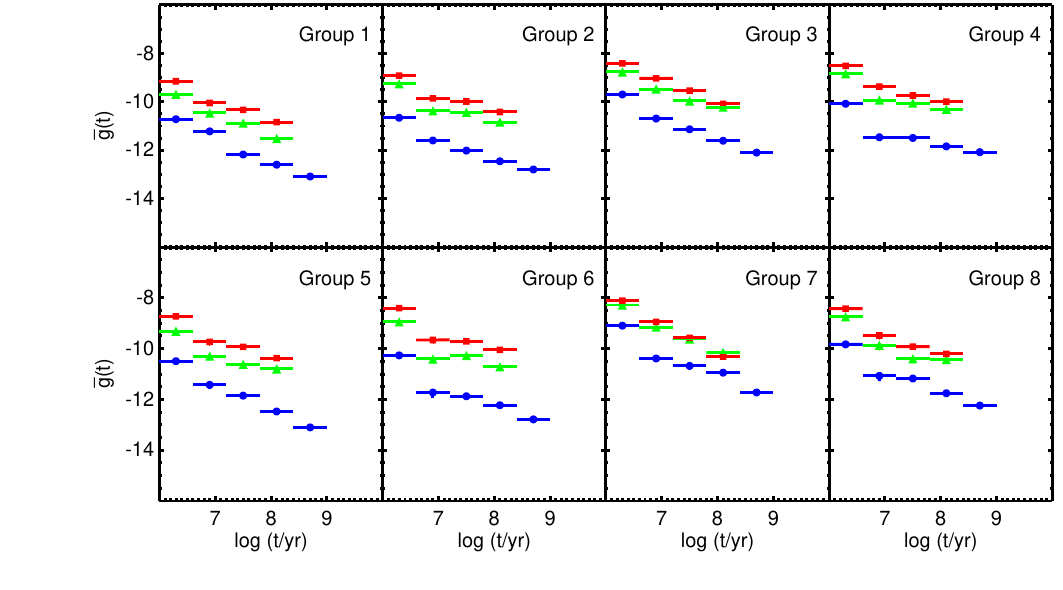}
	\caption{{\bf Age distributions of star clusters for eight galaxy groups have similar declining shapes.}  The groups are arranged in order of decreasing star formation rate starting from the upper left. Each panel shows age distributions in three adjacent mass intervals which stay above the luminosity limit for the galaxies within that group.  The age distributions have been normalized by the mass intervals listed in Table~\ref{tab:gammagroups}, with the highest mass interval in each group plotted in blue, the middle mass interval in green, and the lowest mass interval in red.  The formal uncertainties are smaller than the size of the data points in most cases. All distributions decline steadily starting from the youngest age. There are no statistically significant variations in the shape of the age distribution for different mass clusters in any of the groups, and none across the groups.}.\label{fig-gtgroups}
\end{figure}

Age distributions $\overline{g}(\tau)$ for the galaxy groups, arranged from highest to lowest star formation rate starting from the top-left, are plotted in Figure~\ref{fig-gtgroups}.
In addition to standardizing the age range over which the distributions are constructed, we adopt a uniform set of equal size bins of 0.6 in log~$\tau$, selected to minimize the impact of the well-known age-dating artifacts discussed in Section~3 and still plot at least 4 data points for each distribution. The difference in amplitude between the age distributions in each panel reflects the shape of the mass function of the clusters---since there are more low-mass clusters, the age distributions increase in amplitude with decreasing mass.

\begin{deluxetable*}{lccccccccc}
\tablecaption{Power-law Index $\gamma$ for Age Distribution Fits to Galaxy Groups\label{tab:gammagroups}}
\centering
\tablehead{
\colhead{Group} & SFR & \multicolumn{3}{c}{Mass Range 1} & \multicolumn{2}{c}{Mass Range 2} & \multicolumn{2}{c}{Mass Range 3}   \\ 
 \colhead{\#} & \colhead{($M_{\odot}~\mbox{yr}^{-1}$)} & \colhead{$\gamma_1$ (1~Myr-1~Gyr)} & \colhead{$\gamma_2$ (15~Myr-1~Gyr)} & \colhead{logM} &
 \colhead{$\gamma_3$ (1-250~Myr)} & \colhead{logM} & \colhead{$\gamma_4$ (1-250~Myr)}  & \colhead{logM} 
 }
\startdata
1 & 3.75-12.7  & $-1.02\pm0.08$ & $-0.76\pm0.03$ & $>5.3$	& $-0.98\pm0.07$ & $5.0-5.3$ & $-0.89\pm0.14$ & $4.7-5.0$   \\ 
2 & 3.0-3.3 & $-0.85\pm0.11$ & $-0.65\pm0.05$ & $>5.0$ & $ -0.80\pm0.25$ & $4.7-5.0$ & $-0.77\pm0.19$  & $4.5-4.7$  \\ 
3  & 2.0-2.4 & $-0.95\pm0.10$ & $-0.80\pm0.10$ & $>4.5$ & $-0.81\pm0.12$ & $4.2-4.5$  & $-0.92\pm0.03$ & $4.0-4.2$ \\
4 & 1.0-1.25 & $-0.73\pm0.22$ & $-0.49\pm0.05$ & $>4.5$ & $-0.75\pm0.25$ &  $4.2-4.5$ & $-0.80\pm0.16$ & $4.0-4.2$  \\
5 & 0.8-0.9 & $-1.04\pm0.07$ & $-1.04\pm0.10$ & $>4.5$ & $-0.80\pm0.22$ &  $4.2-4.5$ & $-0.85\pm0.19$ & $4.0-4.2$ \\
6 & 0.6-0.7 & $-0.93\pm0.20$ & $-0.76\pm0.10$ & $>4.5$ & $-0.86\pm0.37$ & $4.2-4.5$ &$-0.83\pm0.30$ & $4.0-4.2$ \\
7 & 0.3-0.4 & $-0.97\pm0.17$ & $-0.87\pm0.25$ & $>4.0$ & $-1.01\pm0.10$ & $3.8-4.0$ & $-1.21\pm0.05$ & $3.6-3.8$ \\
8 & 0.1-0.2 & $-0.92\pm0.15$ & $-0.89\pm0.05$ & $>4.0$ & $-0.93\pm0.29$ & $3.8-4.0$ & $-0.92\pm0.21$ & $3.6-3.8$ \\
\enddata
\end{deluxetable*}

The $\overline{g}(\tau)$ distributions all decline more-or-less continuously with increasing age. They can be approximated by a power law of the form $dN/d\tau \propto \tau^{\gamma}$, determined by fitting the function as $dN/d \mbox{log} \tau \propto \mbox{const}$. Uncertainties for each data point are assumed to be $\pm \sqrt{N}$, where N is the number of clusters in the bin. The best-fit results for $\gamma$ along with the mass intervals used in the fit are recorded in Table~\ref{tab:gammagroups}.
For the age distribution which includes the most massive clusters in each group, we record the results of two fits, one over the full age range from log~$(\tau/{\rm yr})=6.0-9.0$ ($\gamma_1$) and another restricted to ages log~$(\tau/{\rm yr})=7.2-9.0$ ($\gamma_2$) which should include only bound clusters. The age distributions for individual galaxies are constructed using the same method but plotted for only two mass intervals, and are also recorded in the Appendix for both the primary and secondary catalogs. 

The value of $\gamma$ is not significantly affected if we use different bin widths and centers.
While all reported fits to cluster age distributions use the default age bins, we find fits within $\pm0.15$ of those listed in the tables for different binning.
We found nearly identical results as those listed in Column~3 of Table~\ref{tab:gammagroups} if we include all clusters with ages between 7 and 50~Myr in a single bin, an age range affected by systematic uncertainties, as mentioned in Section~2.2.1.
We also check the results from maximum likelihood fits to (unbinned) cluster ages in the same mass-age intervals.
In all cases, these find nearly identical values for $\gamma$ as those listed in Table~\ref{tab:gammagroups}, confirming the steep decline of the cluster age distributions.

\subsection{Results}

Visually, the similarity in the shapes of the cluster age distributions within and between panels in Figure~\ref{fig-gtgroups} is quite striking. None of these distributions are flat, and all decline fairly continuously over the plotted mass-age intervals, which in aggregate cover masses between $4\times10^3~M_{\odot}$ to upwards of $10^6~M_{\odot}$.
The age distributions (Table~\ref{tab:gammagroups}) have best-fit power-law indices $\gamma$ between $-0.49$ and $-1.21$ for the galaxy groups, with a mean and standard deviation from all four fits in eight groups of $\overline{\gamma}=-0.86\pm0.13$. If we consider the shape of the cluster age distribution for the interval with just the most massive clusters in each star-forming group (which allows fits to ages up to 1~Gyr), we find a median power-law index of $\gamma=-0.93$, and a mean and standard deviation of $\gamma = -0.93\pm0.10$ over the full age range.

\begin{figure}[!htbp]
	\centering
\includegraphics[width=9.0cm]{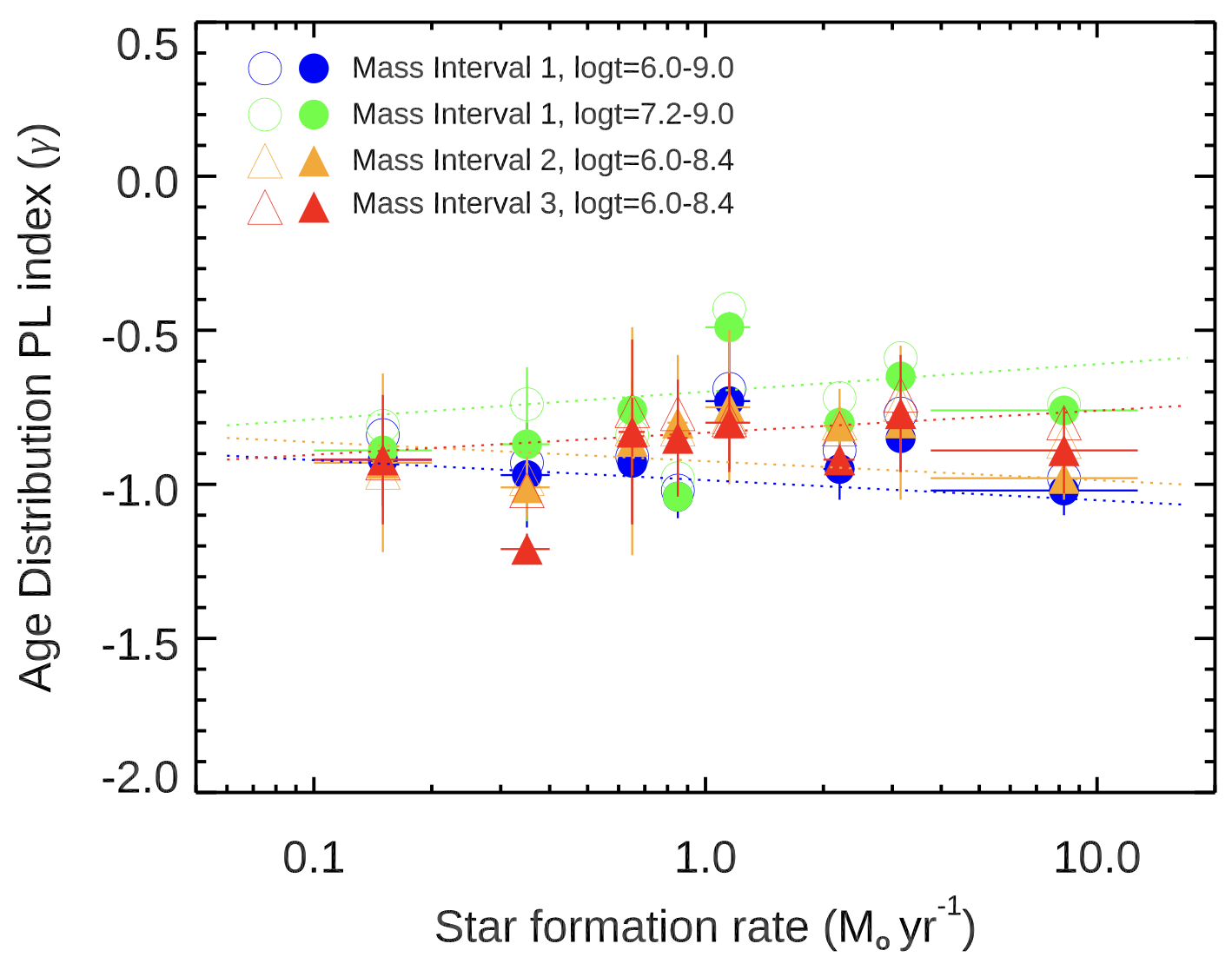}
	\caption{{\bf Shape of cluster age distributions do not depend on SFR of the host galaxy.} Results for the power-law index ($\gamma$) of the cluster age distribution versus SFR.  The filled symbols represent the best fits found for the eight groups based on the primary catalogs. Open symbols show very similar results for the best fits for the secondary catalogs for the same groups and same mass-age limits.}.
    \label{fig-dndt-gamma}
\end{figure}

Figure~\ref{fig-dndt-gamma} shows two of the main results from this work: there is no systematic trend in the steepness of the cluster age distribution $\gamma$ with either cluster mass or the SFR of the host galaxy.
The blue circles represent the fits to $\gamma$ for the highest mass interval in each galaxy group, and are flat across the groups. The results from the machine-learning selected cluster catalogs for the same galaxy groups and age-mass intervals (open blue circles) are similar. The blue, orange, and red points in the figure show that the cluster age distributions in different mass intervals have similar shapes, with a median difference  $\gamma_1$ (blue) minus $\gamma_4$ (red) of 0.08 with a mean and standard deviation of $0.03\pm0.13$. The machine-learning based catalogs have a median difference of 0.07 and mean and standard deviation of $0.05\pm0.14$. We conclude that the age distributions do not depend on cluster mass. 

We estimate the fraction of unbound stellar systems in our sample by comparing the power-law index measured for the age distribution with and without sources younger than 15~Myr (blue vs.\ green points). The distributions all show a steady decline between 15~Myr and 1~Gyr (the power-law indices are recorded in column~4 of Table~\ref{tab:gammagroups}), that are somewhat shallower than the fit over the entire 1~Myr to 1~Gyr age range, with a mean and standard deviation of $-0.78 \pm 0.16$ for the former and  $-0.93\pm0.10$ for the latter. The difference in the power-law indices, $\gamma_2 - \gamma_1$, ranges from 0.0 to 0.26 for the galaxy groups, with a median difference of 0.17. For machine-learning selected cluster catalogs $\gamma_2 - \gamma_1$ ranges from 0.03 to 0.24, with a median difference of 0.18. If we attribute this difference to unbound stellar systems, it implies that $\approx30$\% $(1-\frac{1}{10^{0.17}}$) of sources younger than 15~Myr in our catalogs are unbound. 

\begin{figure}[!htbp]
	\centering
\includegraphics[width=\textwidth]{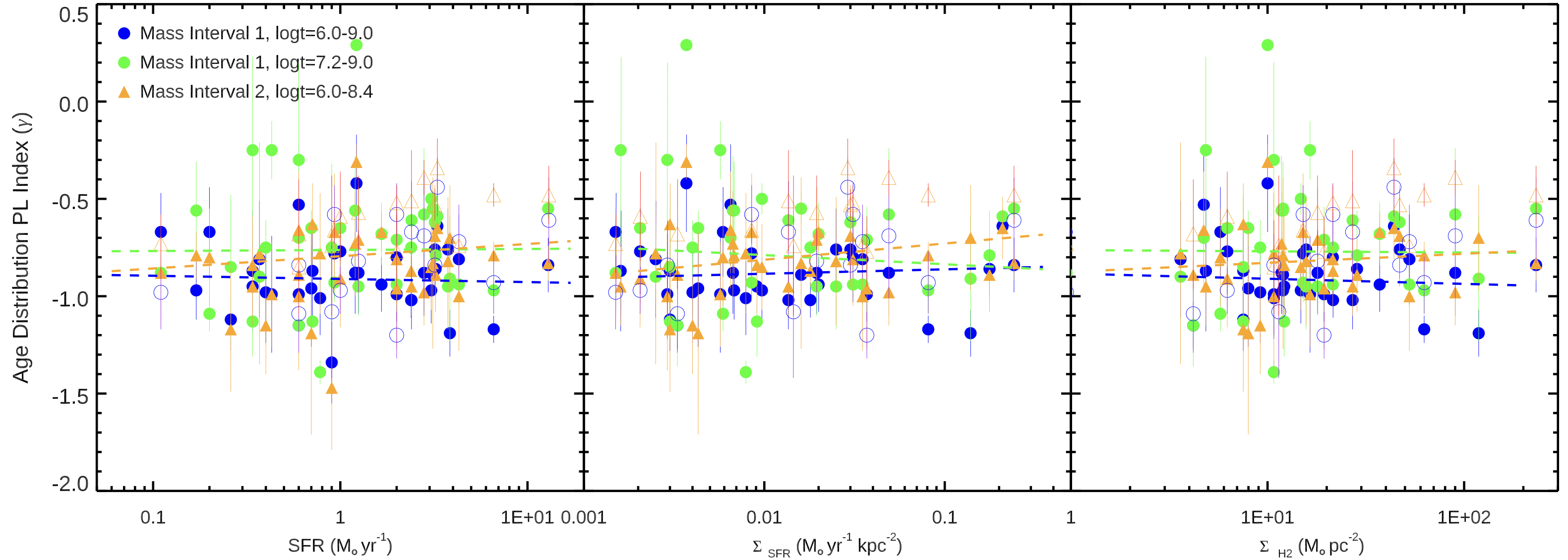}
	\caption{{\bf There is no statistically significant trend between the power-law index $\gamma$ measured for age distributions in individual galaxies and cluster mass or host galaxy properties. } The best fit power-law index $\gamma$ to the age distributions of all 38 galaxies are plotted against the SFR, $\Sigma_{\rm SFR}$, and $\Sigma_{\rm H_2}$ for the host galaxy.  The mass intervals used for each distribution are recorded in Table~\ref{tab:gammagals}, and the legend indicates the age range of the clusters included in the fit.  Solid circles show the results from the primary catalogs, while open circles represent the best fits to the catalogs of \citet{Henny25} for 16 galaxies. The shape of the age distribution is similar for clusters in a range of mass intervals. }.
    \label{fig-gammagals}
\end{figure}

Results for $\gamma$ in all 38 galaxies individually are shown in Figure~\ref{fig-gammagals} versus SFR (left), $\Sigma_{\rm SFR}$ (middle), and $\Sigma_{\rm H_2}$ (right).
We plot two mass intervals per galaxy and ensure the higher mass interval is complete to log~$(\tau/{\rm yr})=9.0$ and the lower mass interval is complete to log~$(\tau/{\rm yr})=8.4$. The higher mass interval is again fit over two age ranges, one between log~$(\tau/{\rm yr})=6.0-9.0$ (blue) while the other excludes the youngest objects and fits clusters between log~$(\tau/{\rm yr})=7.3-9.0$ (green), as indicated in the legend.  The orange points show $\gamma$ results for the lower mass interval in each galaxy (listed in Column~6 of Table~\ref{tab:gammagals}). The open symbols represent the best-fit to  the \citet{Henny25} cluster catalogs (listed in Table~\ref{tab:gammahenny}). The results for the individual galaxies (from both sets of age-dating) are mostly between $\gamma \approx -0.5$ and $-1.0$ regardless of age and mass interval, similar to those found for the composite groups. If we consider the best-fit value of $\gamma$ for all three of the power-law fits listed in Table~\ref{tab:gammagals}, {\em we find a mean and standard deviation of $\overline{\gamma}=-0.83\pm0.24$}; we adopt this value to characterize our sample. No statistically significant trend between $\gamma$ and SFR, $\Sigma_{\rm SFR}$ or the surface mass density of molecular gas is found, since all fits in Figure~\ref{fig-gammagals} (represented by the dashed lines) are consistent with a slope of 0.

We conclude that empirical age distributions of star clusters in spiral galaxies have similar declining shapes regardless of the details of source selection, age-dating, and the mass-age intervals used to construct the distributions. In all cases, the cluster age distributions decline steeply and more-or-less continuously over the past $\approx1$~Gyr, with a power-law index $\gamma=-0.83\pm0.24$, and the distributions are independent of cluster mass and host galaxy SFR, $\Sigma_{\rm SFR}$, and $\Sigma_{\rm H_2}$.

\section{Discussion}\label{sec-discussion}

\subsection{Cluster Populations in Spirals are Dominated by Mass-Independent Dissolution over first $\approx1$~Gyr}

Two of the main results from Sections~\ref{sec-cmf} and \ref{sec-dndt} are that cluster mass and age distributions aggregated across all environments in spiral galaxies are independent of one another for the first $\approx1$~Gyr. This means the joint distribution of cluster masses and ages can be written as a product:
\begin{equation}
    g(M,\tau) \propto M^{\beta} \tau^{\gamma}.
\end{equation}
We found a power-law index of $\beta=-1.9 \pm0.2$ for the mass function and $\gamma=-0.83\pm0.24$ for the age distribution describe the cluster populations in our spiral sample. These results, based on thousands of star clusters in 38 spiral galaxies, are independent of the specific cluster selection procedure, age-dating method, or age-mass intervals. 
Similar trends are recovered independently from the human- and machine-learning-selected catalogs, from two different age-dating approaches, and from both the individual galaxies and the composite galaxy groups. The consistency across these different analyses argues that the observed similarities in the mass and age distributions are not primarily driven by selection effects or methodological uncertainties.

Observed age distributions reflect the difference between the formation and dissolution rates of the clusters, $\gamma = \gamma_{\rm form} - \gamma_{\rm disrupt}$. Spiral galaxies in general \citep[][]{Lee09,SilvaVilla11,Lewis15,Sacchi19,Lazzarini22} and certainly the combined groups studied here are expected to have experienced fairly constant star formation histories over the past Gyr, so we can assume that $\gamma_{\rm form} \approx0$.\footnote{The PHANGS collaboration will independently determine the star formation histories of the galaxies from spectroscopy taken as part of PHANGS-MUSE \citep{Emsellem2022}.} The steep decline observed in the age distributions therefore indicates strong, early cluster dissolution. This is true when considering both the age and mass distributions of bound clusters, which tell a consistent story, with $\gamma \approx -0.8$ for cluster age distributions between 15~Myr and 1~Gyr (see Section~5), and a corresponding decline in amplitude of the mass functions between $10-100$~Myr and $0.4-1$~Gyr. 

If the observed cluster age distributions were driven primarily by variations in cluster formation history rather than by dissolution, the implied cluster formation rates would need to rise steeply toward the present day. For a power-law index of $\gamma=-0.83$, a pure formation-history interpretation would correspond to an increase in the cluster formation rate by a factor of $\sim300$ over the past Gyr $(10^3)^{0.83}\sim300$. Resolved-stellar-population studies of nearby spiral galaxies generally do not support such large increases in star formation activity \citep[e.g.,][]{Lee09,SilvaVilla11,Lewis15,Sacchi18,Lazzarini22}, instead finding variations that are typically much more modest over the past Gyr. While variations in star formation history likely contribute some scatter in the observed age distributions, it is difficult to explain the similar slopes observed across 38 galaxies and the composite galaxy groups as arising primarily from coordinated increases in cluster formation histories. We therefore conclude that cluster dissolution must play a major role in shaping the observed age distributions.

We conclude that empirical age and mass distributions indicate that cluster populations in spirals evolve significantly over their first Gyr and that their demographics are dominated by dissolution (rather than by formation) that does not depend on the initial mass of the clusters, or {\em mass-independent dissolution}. A similar conclusion was reached for dwarf galaxies studied as part of the LEGUS project \citep{Calzetti2015}.
As found here for massive spirals, the shapes of the cluster mass and age distributions in these lower-mass galaxies do not appear to correlate with SFR or $\Sigma_{\rm SFR}$ \citep[]{cook19,Cook23}, and their combined star formation histories are approximately constant over the last few hundred Myr (from applying a synthetic color-magnitude method to hundreds of thousands of individual stars; \citealt{cignoni19}). 

\begin{figure}[!htbp]
    \centering
    \includegraphics[width=7.5cm]{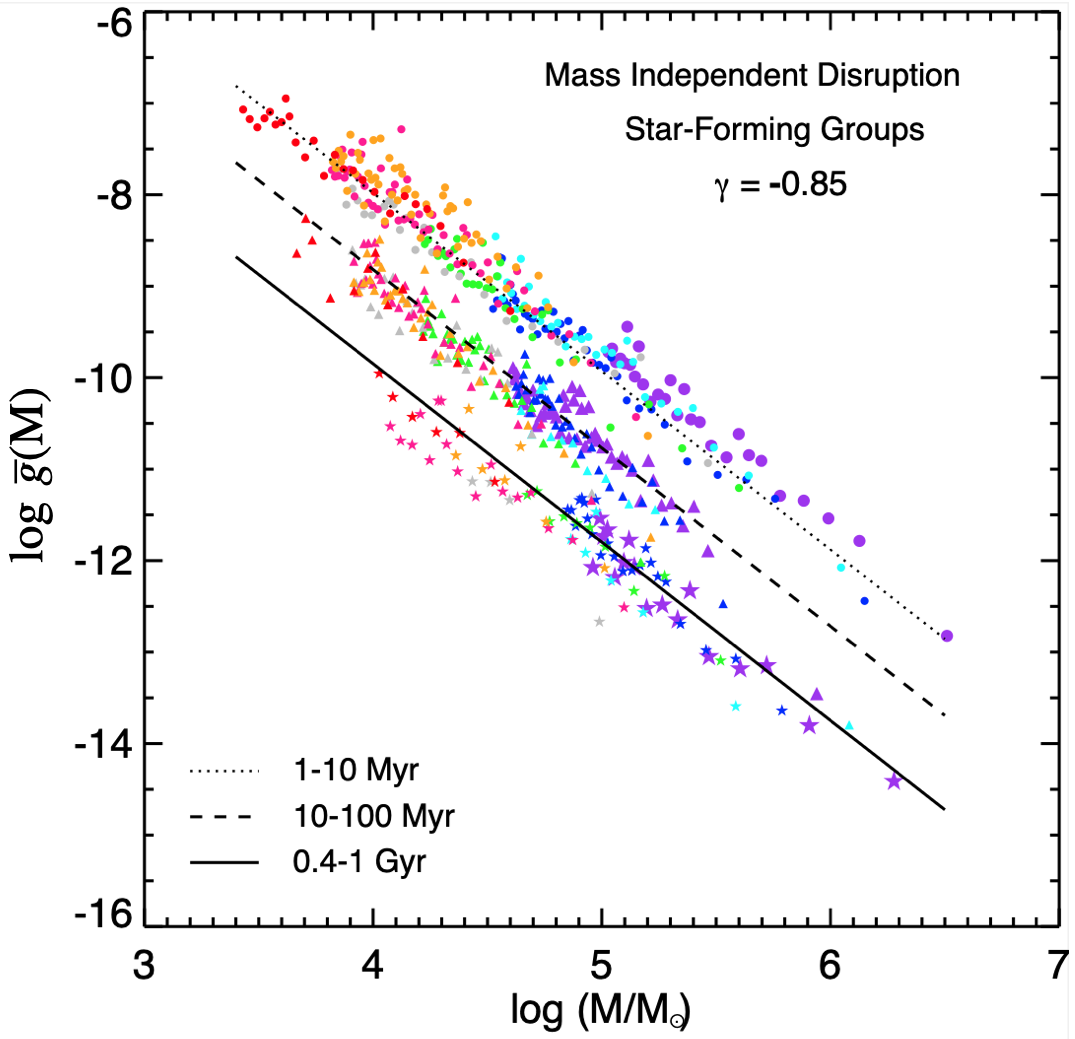}
    \includegraphics[width=7.5cm]{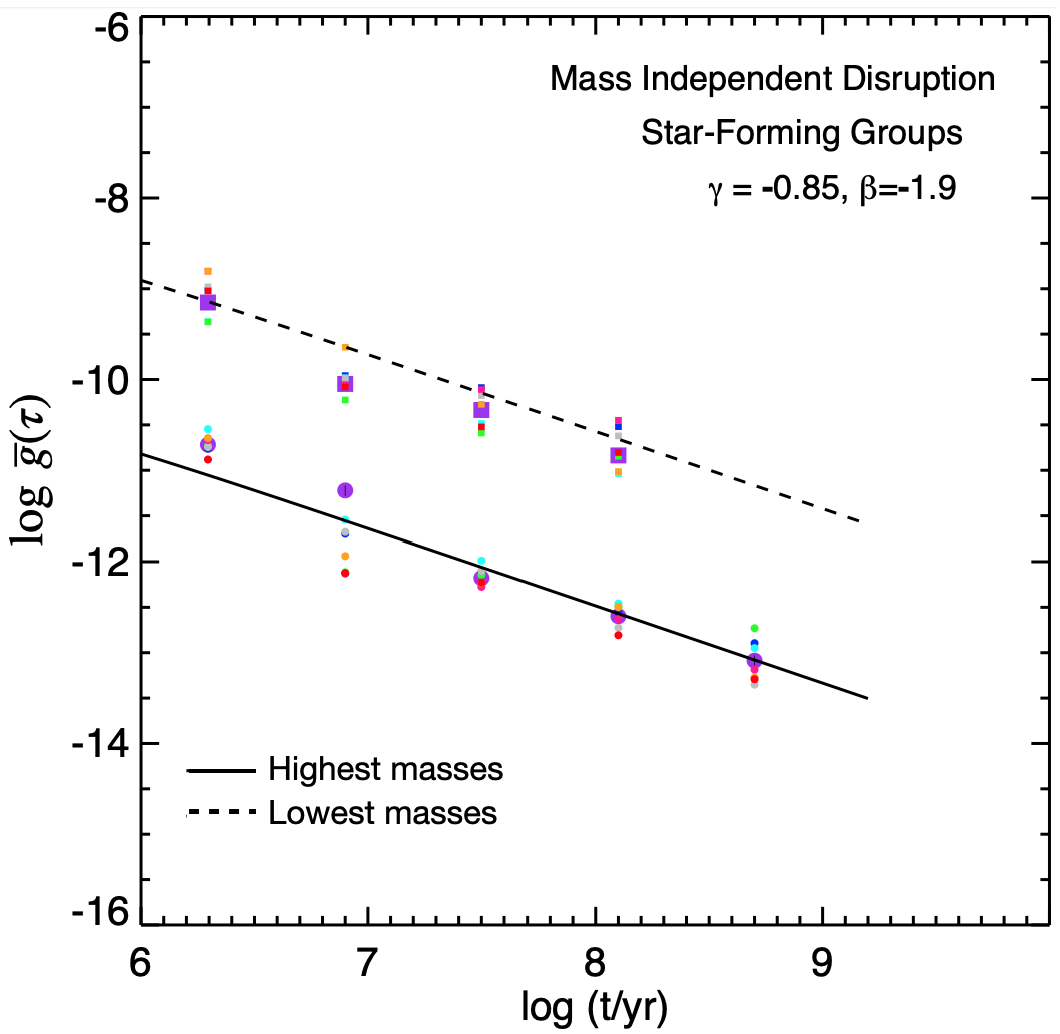}
	\caption{{\bf Mass-independent dissolution does not alter the shapes of the mass or age distributions, but decreases the amplitude of the mass function over time.} The left panel plots together the mass functions of all 8 galaxy groups, where the amplitudes of the observed mass functions are normalized only so that $1-10$~Myr clusters from groups 2-8 match those of group 1, and the differences in amplitude between the $10-100$~Myr and $0.4-1$~Gyr distributions are preserved between each galaxy group. Together, the galaxy groups form a nearly seamless power-law distribution covering a factor of $\approx1000$ in cluster mass for all three age intervals.  The black lines in the left panel show the predicted evolution of the cluster mass function $\overline{g}(M)$ from a mass-independent dissolution model with $\gamma = -0.85$. The right panel plots together the age distributions from all 8 galaxy groups, where groups 2-8 are normalized to match group 1. The age distributions have similar shapes even though a different mass interval was used for each group. The predicted shape of the age distribution is similar for different mass intervals when cluster dissolution is independent of mass.     
     See section~6.1 for details. }.\label{fig-modelMID}
\end{figure}

In the left panel of Figure~\ref{fig-modelMID} we plot the observed cluster mass functions from Section~4 for the $1-10$~Myr, $10-100$~Myr, and $0.4-1$~Gyr age intervals for all eight galaxy groups.
The different color data points represent the different groups, with the mass function for group 1 shown in purple (highest SFR) and for group 8 in red (lowest SFR). The normalization between the distributions in the three age intervals has been preserved for each group, with the $1-10$~Myr mass functions for groups 2-8 shifted up vertically to match that for group 1. Plotted together, it is immediately clear that the `initial' (1-10~Myr) mass function of star clusters in spiral galaxies (top set of data points) have a remarkably consistent shape over a factor of $\approx1000$ in mass, and form essentially a single continuous power-law distribution over the mass range log~$(M/M_{\odot})=3.5-6.5$. The combined mass function for the other two plotted age intervals, $10-100$~Myr and $0.4-1$~Gyr, also form essentially a single power-law with a drop in amplitude that is remarkably similar across all of the groups.

The right panel of Figure~\ref{fig-modelMID} plots the observed age distributions of star clusters in all eight composite groups, with the same color coding as the left panel. Here we show the distributions observed for the highest and lowest mass intervals, which are the lower and upper distributions respectively, and are different for each star-forming group. The age distributions for star-forming groups 2-8 have been shifted vertically to match that of group~1. Overall, the age distributions of clusters in galaxies with different star-forming properties show a similar steep decline, with no obvious difference in the shape of the distributions between the two mass intervals. Note that the decline is not driven by sources younger than 10~Myr.

\begin{figure}[!htbp]
	\centering
\includegraphics[width=12cm]{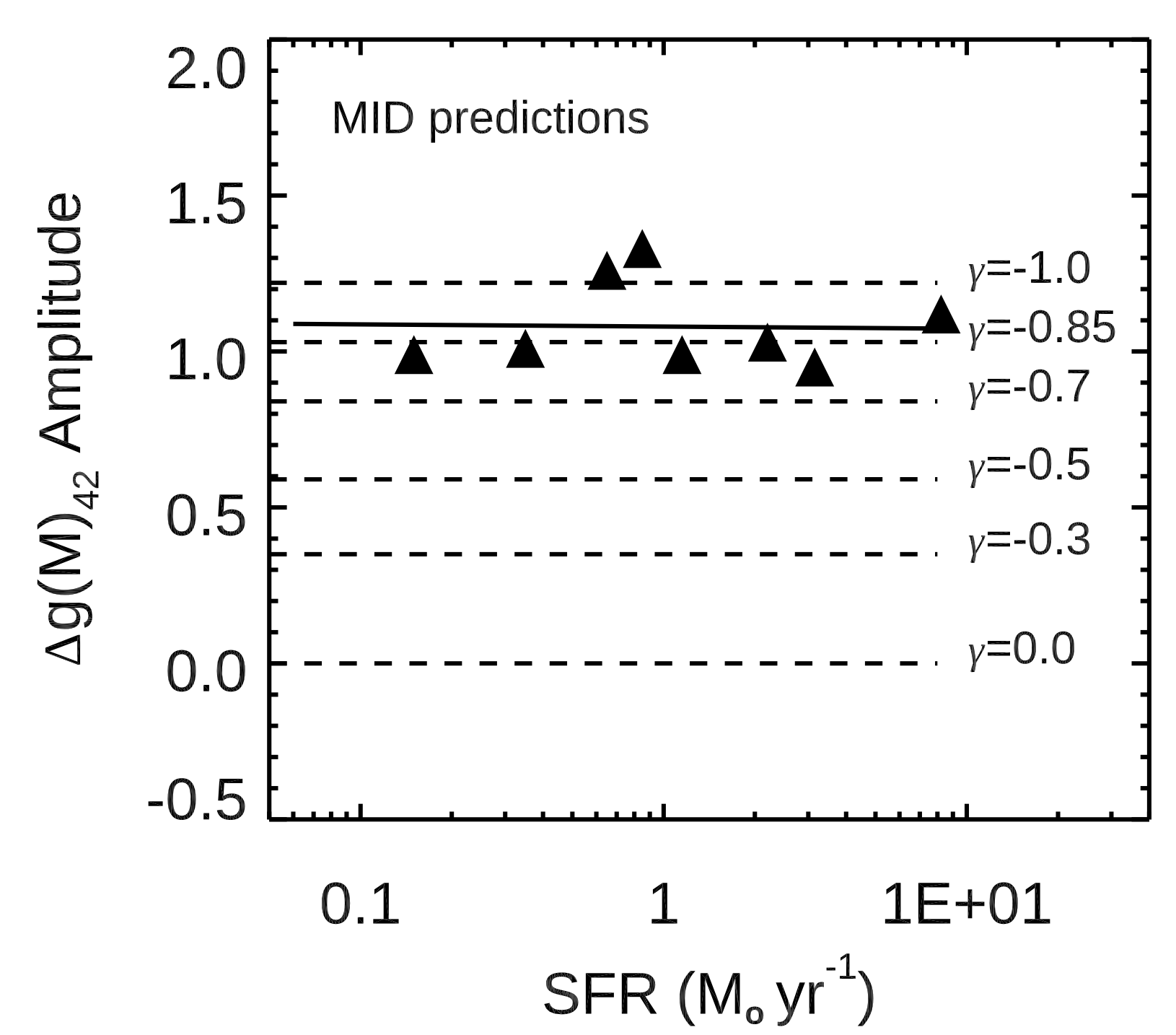}
	\caption{{\bf The decrease in the amplitude of the cluster mass function does not vary with properties of the host galaxy.} The drop in amplitude between $10-100$~Myr and $0.4-1$~Gyr clusters, $\Delta~g_{42}$ is best matched by the mass-independent disruption model with a power-law index for the age distribution of $\gamma \approx-0.85$.  The dashed lines show the predicted decrease in amplitude between clusters in these two age intervals for different rates of mass-independent disruption. }.\label{fig-Deltag}
\end{figure}

Predictions from the mass-independent dissolution model presented by \citet{fall09} for $\beta=-1.9$ and $\gamma=-0.85$ are represented by the lines in each panel of Figure~\ref{fig-modelMID}. The model predictions are normalized to the youngest age interval for the mass function (left panel) and to the highest mass interval for the age distribution (right), represented by the filled circles and dotted line. In the left panel, the dashed and solid lines show predictions for $10-100$~Myr and $0.4-1$~Gyr clusters, respectively. These predictions are well-matched to both the shape and amplitude of the observed mass functions, and hence to the evolution of the cluster population. Figure~\ref{fig-Deltag} compares the observed drop in amplitude between the cluster mass function for $10-100$~Myr and $0.4-1$~Gyr ($\Delta gM_{42}$) with different rates of mass-independent disruption, with $\gamma$ values of 0.0, $-0.3$, $-0.5$, $-0.7$, $-0.85$, and $-1.0$.
The value of $\gamma=-0.85$ clearly gives a good representation of $\Delta g_{42}$ measured for the galaxy groups, while values less than this predict a smaller drop in amplitude than we observe.

The mass-independent dissolution model from \citet{fall09} describes the situation where entire clusters gradually dissolve independent of their initial mass. Since the timescale for dissolution depends on cluster density which is expected to be fairly constant in galaxies due to tidal limitation, it is reasonable that dissolution times do not depend strongly on cluster mass (see discussion in Section~6.3). A value of $\gamma_{\rm disrupt}=-0.85$ implies that $\approx85$\% ($1 - (10^{-0.85})^{-1}$)
of clusters dissolve for each factor of 10 increase in age. This type of evolution also describes different combinations of survival probability and mass-loss, where some clusters lose a fraction of their mass while others dissolve completely.

While a near-constant rate of mass-independent dissolution drives the observed mass-age demographics of star cluster populations on galaxy scales, significant variations are observed for mass-age distributions on $\lea0.5$~kpc physical scales \citep[e.g.,][]{bastian11,Fouesneau14,Messa18,Linden22}.
Smaller regions can experience strong variations in their star formation histories, with older clusters typically born elsewhere.
As many small regions are combined into larger ones, differences in star formation histories average out.

\subsection{Constraints on Mass-Dependent Dissolution}

While dissolution processes that operate independent of mass appear to dominate cluster demographics over the first Gyr or so, mass-dependent dissolution processes, particularly stellar evaporation due to relaxation, are expected to play an increasingly important role on longer timescales. Interactions between cluster stars drive them toward energy equipartition, resulting in lower mass stars with sufficiently high velocities that they can escape the cluster \citep[e.g.,][]{Henon61,Fukushige00,baumgardt03}. Relaxation causes a power-law mass distribution to evolve into a peaked one after a few billion years, as observed for ancient globular cluster populations in spiral galaxies \citep[e.g.,][]{Fall01,Chandar07,McLaughlin08}.

\begin{figure}[!htbp]
	\centering
       \centering
    \includegraphics[width=7.5cm]{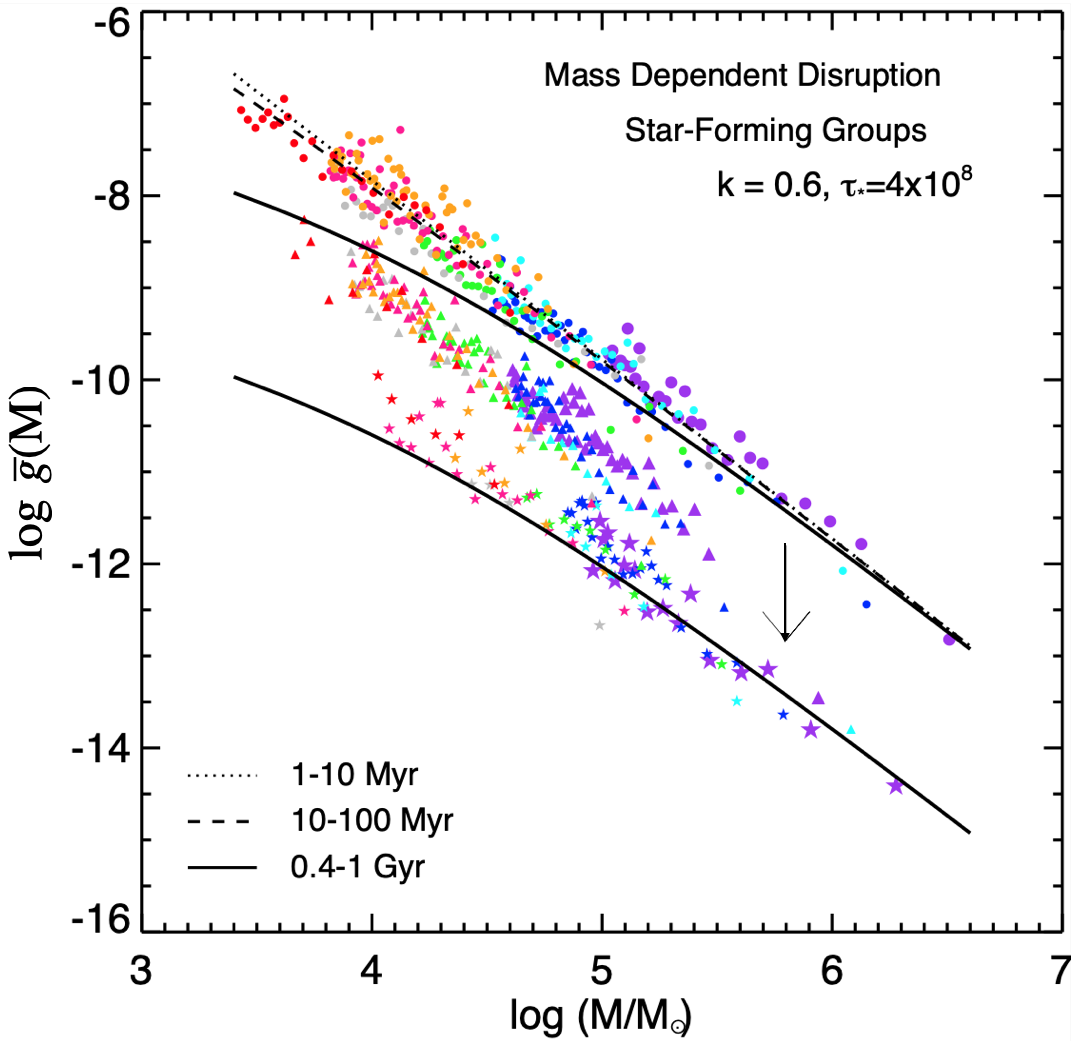}
    \includegraphics[width=7.5cm]{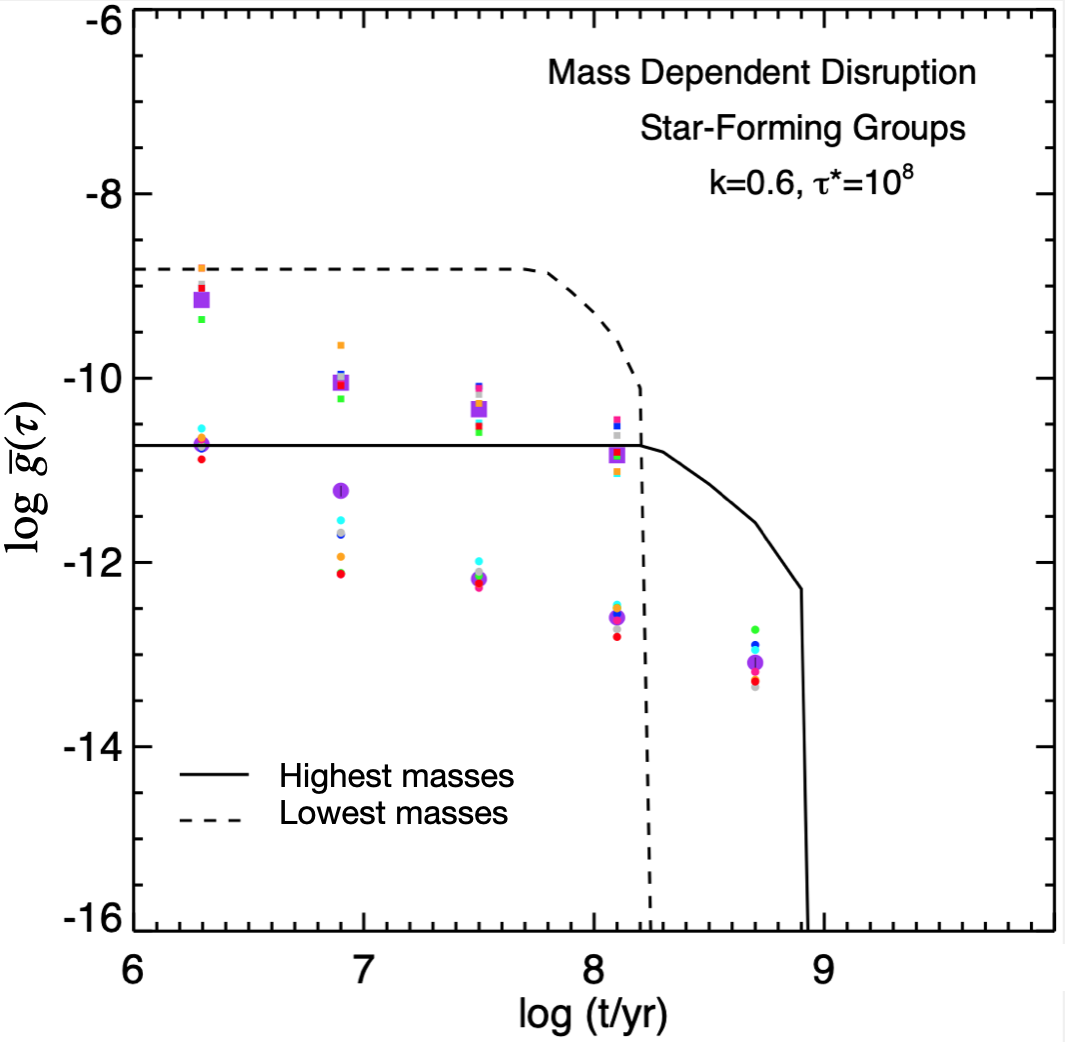}
	\caption{{\bf Mass-dependent dissolution alters the shape of the mass function at older ages and the age distribution at lower masses.} Predictions for the evolution of the cluster mass (left) and age distributions (right) for a disruption index $k=0.6$ and a characteristic disruption times $\tau_*= 4\times10^8$~yr are shown.
    The dotted line in the left panel is the `initial' mass function (1-10~Myr) clusters, while the solid black curve is the predicted shape for $0.4-1$~Gyr clusters for the given $k$ and $\tau_*$ values. Shorter disruption times and higher $k$ values lead to stronger flattening.  The age distributions are predicted to be flat until they drop exponentially, with the distribution for lower-mass clusters dropping earlier than for higher-mass ones.}\label{fig-modelMDD}
\end{figure}

Figure~\ref{fig-modelMDD} illustrates how the mass (left) and age (right) distributions would evolve if mass-dependent dissolution dominated the first Gyr.
The same empirical distributions from the eight groups presented in Figure~\ref{fig-modelMID} are reproduced here. The lines in each panel illustrate the predicted evolution of the distributions for a specific mass-dependent dissolution model. Mass-dependent dissolution models can be parametrized as:
\begin{eqnarray}
    dM/d\tau &=& -M/\tau_d(M),\\
    \tau_d(M)&=&\tau_*(M/M_*)^k,
\end{eqnarray}
where the exponent $k$ is the disruption index and $\tau_*$ is the characteristic disruption timescale for a $10^4~M_{\odot}$ cluster. Both $k$ and $\tau_*$ are adjustable parameters in the model (see \citealt{fall09} for more details).

As illustrated below, when dissolution depends on cluster mass, the shapes of the mass and age distributions change over time and become coupled. In the left panel of Figure~\ref{fig-modelMDD} an initial cluster mass function with $\beta=-1.9$ (dotted line) is normalized to the observed $1-10$~Myr distribution. The solid line is the mass function predicted for 0.4--1~Gyr clusters calculated for the specific parameters $k=0.6$ and $\tau_*= 5\times10^8$~yr, where we have assumed a constant formation rate. Higher (lower) values of $k$ lead to stronger (weaker) curvature towards lower masses at a fixed $\tau_*$, and shorter (longer) characteristic dissolution timescales $\tau_*$ lead to stronger (weaker) predicted curvature at a fixed value of $k$. The predicted amplitude is quite similar, only the {\em shape} of the distribution changes (flattens) towards lower masses at older ages.
Clearly these models cannot reproduce the change in amplitude of the cluster mass function over time. 

To compare the predicted and observed {\em shapes}, we offset the predicted $0.4-1$~Gyr solid curve vertically down to match the high-mass end of the observed distribution (denoted by the arrow). The particular combination of $k$ and $\tau_*$ shown in the figure predicts that the mass function of $0.4-1$~Gyr clusters should begin to flatten and deviate away from a power-law at masses near log~$(M/M_{\odot})=4$. A qualitative comparison between the observed distribution and predictions from the mass-dependent dissolution model suggests the characteristic disruption timescale for a $10^4~M_{\odot}$ cluster must be longer than $> 4\times10^8$~yr for both $k=0.6$ (shown) and $k=1.0$ (not shown). This lower-limit for $\tau_*$ is consistent with the expectations of relaxation-driven evaporation shaping the observed mass function of globular clusters after $\approx10$~Gyr of evolution from an initial power-law into a log-normal distribution with a peak near $\approx10^5~M_{\odot}$.

The predicted age distributions for $k=0.6$ and $\tau_*=4\times10^8$~yr also do not match the observed cluster distributions in the right panel of Figure~\ref{fig-modelMDD}.
They are predicted to be flat starting at log~$(\tau/\mbox{yr})=6.0$ for a constant star formation rate, then to drop exponentially at an age that depends on the specific mass interval (calculated here for the highest and lowest mass intervals used for galaxy group 1).
The age distribution for the lower-mass clusters is predicted to drop earlier than for the higher mass clusters, since lower-mass clusters (at a given density) dissolve faster than higher-mass ones in mass-dependent dissolution models.

Empirical mass and age distributions of open clusters identified in the solar neighborhood show clear signatures of mass-dependent dissolution, with flatter mass functions observed for older clusters and steeper age distributions for lower-mass clusters \citep{Hunt24}. These Milky Way open clusters however have significantly lower masses than the samples studied here, extending to $\approx60~M_{\odot}$ with only a handful more massive than $10^3~M_{\odot}$.

\subsection{Constraints on Cluster Dissolution by Molecular Structures in Spiral Disks}

The raw material necessary for formation also helps to destroy previously formed clusters. As clusters orbit in the disk of their host galaxy they experience gravitational shocks as they pass by molecular structures, which increases their internal energy and leads to stellar mass-loss and dissolution \citep[e.g.,][]{Spitzer58}. This dissolution mechanism is not mass-dependent, i.e.\ it will not change the observed shape of the cluster mass function, when the half-mass density, $\rho_{\rm h}$, and mass of clusters are uncorrelated, since the dissolution timescale depends on the properties of clusters only through $\rho_{\rm h}$ \citep[e.g.,][]{Binney08,fall12,Krumholz19}. The half-mass density $\rho_{\rm h}$ is expected to be uncorrelated with mass if clusters are tidally limited, since the tidal field of the host galaxy tends to impose the same mean density on clusters at the same galactocentric distance, independent of their mass \citep[e.g.,][]{King62,Innanen83,fall12}.

Measuring the structure of clusters at the distances of PHANGS galaxies is challenging. While it is not possible to measure their core and tidal radii directly, it is possible to fit an assumed functional form (convolved with the point-spread function) and estimate the projected effective radius, $r_e$. 
The PHANGS collaboration is estimating $r_e$ using a custom-built fitting pipeline that minimizes the difference between $30\times30$ pixel cutouts around a cluster and the ePSF-convolved model image. To produce a separate ePSF for each field, we manually select 10 bright isolated stars in each galaxy with the lowest background level and and use the \texttt{EPSFBuilder} routine of Astropy package \texttt{Photutils} \citep{photutils}. We fit the cluster light with the EFF surface brightness profile, which is appropriate for young star clusters \citep[][]{elson87, larsen99, McLaughlin05}. We generalize the EFF profile to non-circular clusters, with a variable axis ratio and position angle, and allow the fit to iteratively find a true cluster center.  To account for contaminating sources inside cluster snapshots, we mask all point sources and any pixels associated with another source that is within 6 pixels of a star cluster.
This automated masking produces successful fits even when there are other sources nearby. We limit the model profile to the maximum radius of the snapshot, which enables reliable effective radius measurements in cases where the surface density profile is very shallow.

\begin{figure}[!htbp]
	\centering
\includegraphics[width=16.0cm]{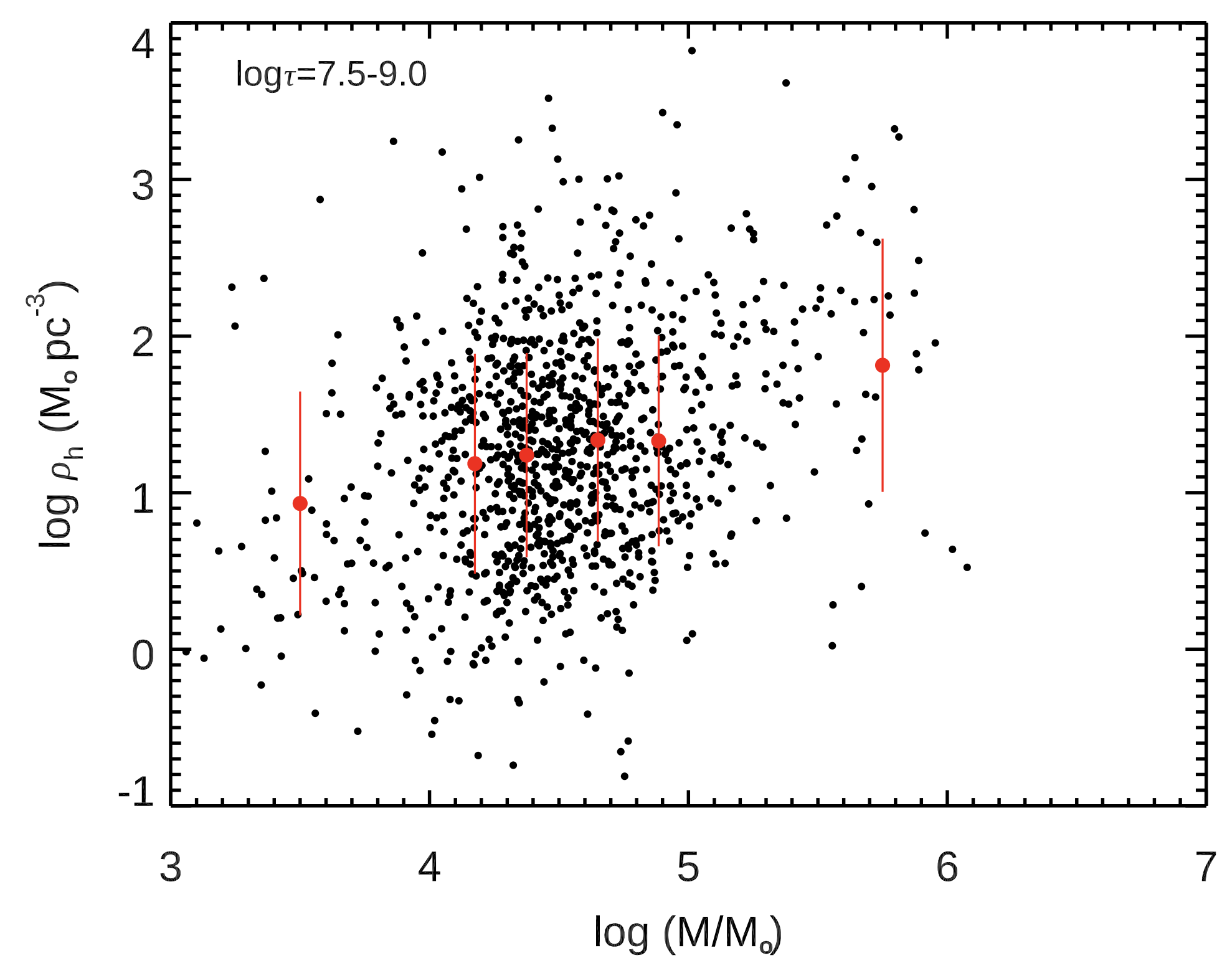}    
	\caption{{\bf Preliminary estimates of stellar half-mass density $\rho_{\rm h}$ show little-to-no correlation with cluster mass.}  The sample is restricted to clusters with estimated ages between $3\times10^7-10^9$~yr in 12 PHANGS galaxies, in order to exclude unbound sources.  The red points are the median and standard deviation in different mass intervals.}.\label{fig-rhovsmass}
\end{figure}

For all clusters in the PHANGS-HST sample that currently have both mass and preliminary $r_{\rm e}$ estimates available, we calculate the internal half-mass density $\rho_{\rm h}$ (the mean density within the half-mass radius $r_{\rm h}$) from the definition $\rho_{\rm h} \equiv 3M/8\pi r_{\rm h}^3$, where we have assumed the 3D-2D conversion $r_{\rm h}=(4/3)r_{\rm e}$ \citep{Spitzer1978}. We plot the relationship between $\rho_{\rm h}$ and mass for bound clusters by restricting the sample to ages between 30~Myr and 1~Gyr. Figure~\ref{fig-rhovsmass} shows there is a lot of scatter between $\rho_{\rm h}$ and mass.
The median $\rho_{\rm h}$ and standard deviation for different mass intervals are plotted as the red points and bars. Although not definitive, this preliminary plot suggests a lack of correlation between $\rho_{\rm h}$ and mass, consistent with recent findings for the mass-radius relationship measured for clusters in 31 nearby star-forming galaxies observed as part of the LEGUS survey \citep{Brown21} and in the disk of M82 \citep{Cuevas21}, as expected for mass-independent dissolution.

While dissolution does not appear to alter the shape of the mass function as clusters evolve, it is plausible that the {\em rate} of mass-independent dissolution could depend on the amount of molecular material in the galaxy \citep[e.g.,][]{Boutloukos03}. In this case, galaxies with higher gas surface densities $\Sigma_{\rm H_2}$ should experience higher rates of mass-independent dissolution, resulting in steeper age distributions and larger decreases in amplitude between the mass functions of $10-100$~Myr and $0.4-1$~Gyr clusters. Our empirical results do not reveal such trends. This lack of correlation between the age-distribution index $\gamma$ and $\Sigma_{\rm H_2}$ and between the decrease in the amplitude of the mass function $\Delta gM_{42}$ and $\Sigma_{\rm H_2}$ may indicate that the factor of $\approx100$ range in average gas surface density is insufficient to drive different dissolution rates. Said another way, our study suggests that the smallest surface density of molecular material ($\approx1~M_{\odot}~\mbox{pc}^{-2}$) in our sample is sufficient to disrupt clusters at a rate similar to that in galaxies with the highest surface density ($\approx200~M_{\odot}~\mbox{pc}^{-2}$).

\subsection{Implications for Cluster Formation and Evolution in Early Galaxies}

Observations with JWST have unexpectedly revealed compact star clusters forming in galaxies at high redshift, when the universe was only $\approx0.5$~Gyr old. 
Gravitational lensing gives an exceptionally sharp view of bright, compact star clusters in the Sparkler ($z=1.37$; \citealt{Mowla22,Adamo23,Tomasetti25}), the Relic ($z=2.35$; \citealt{Whitaker25}), Sunburst Arc ($z=2.37$; \citealt{Vanzella22}), 
Sunrise Arc ($z\sim6$; \citealt{Vanzella23}), and Cosmic Gems Arc ($z=9.6$; \citealt{Adamo24}).
Young clusters in these early galaxies are forming in the disks (not halos) of their parent galaxies \citep[e.g.,][]{Vanzella22}, and have half-light radii or r$_{\rm eff}$ between $\approx1$ and 10~pc \citep[e.g.,][]{Vanzella22,Vanzella23,Adamo24}, similar to those of young clusters in nearby spiral galaxies.
Studies suggest the presence of such massive clusters points to unexpectedly high rates of star formation in the early universe.

The estimated ages and masses for star clusters at high redshift have some overlap with those in our spiral sample. The brightest cluster in the Sunburst arc is $\approx3$~Myr old, has r$_{\rm eff} \approx 8$~pc, and a dynamical and stellar mass $\sim10^7~M_{\odot}$ \citep{Vanzella22}.  The system has formed at least 12 additional detectable clusters with estimated masses of a few$\times10^6~M_{\odot}$. The more distant Sunrise arc has six detected compact young clusters, including a few $\lea5$~Myr line-emitting sources, with estimated masses in the range $\approx10^6-10^7~M_{\odot}$ \citep{Vanzella23}. Five compact (r$_{\rm eff} \approx 1$~pc), young ($<50$~Myr) clusters have been discovered in the Cosmic Gems Arc (viewed just 450~Myr after the Big Bang), and have estimated masses $\approx10^6~M_{\odot}$. As discussed in Section~3, nearby spiral galaxies NGC~1365 and NGC~1672 are currently forming clusters with masses higher than $10^6~M_{\odot}$.

\begin{figure}
	\centering
\includegraphics[width=16.0cm]{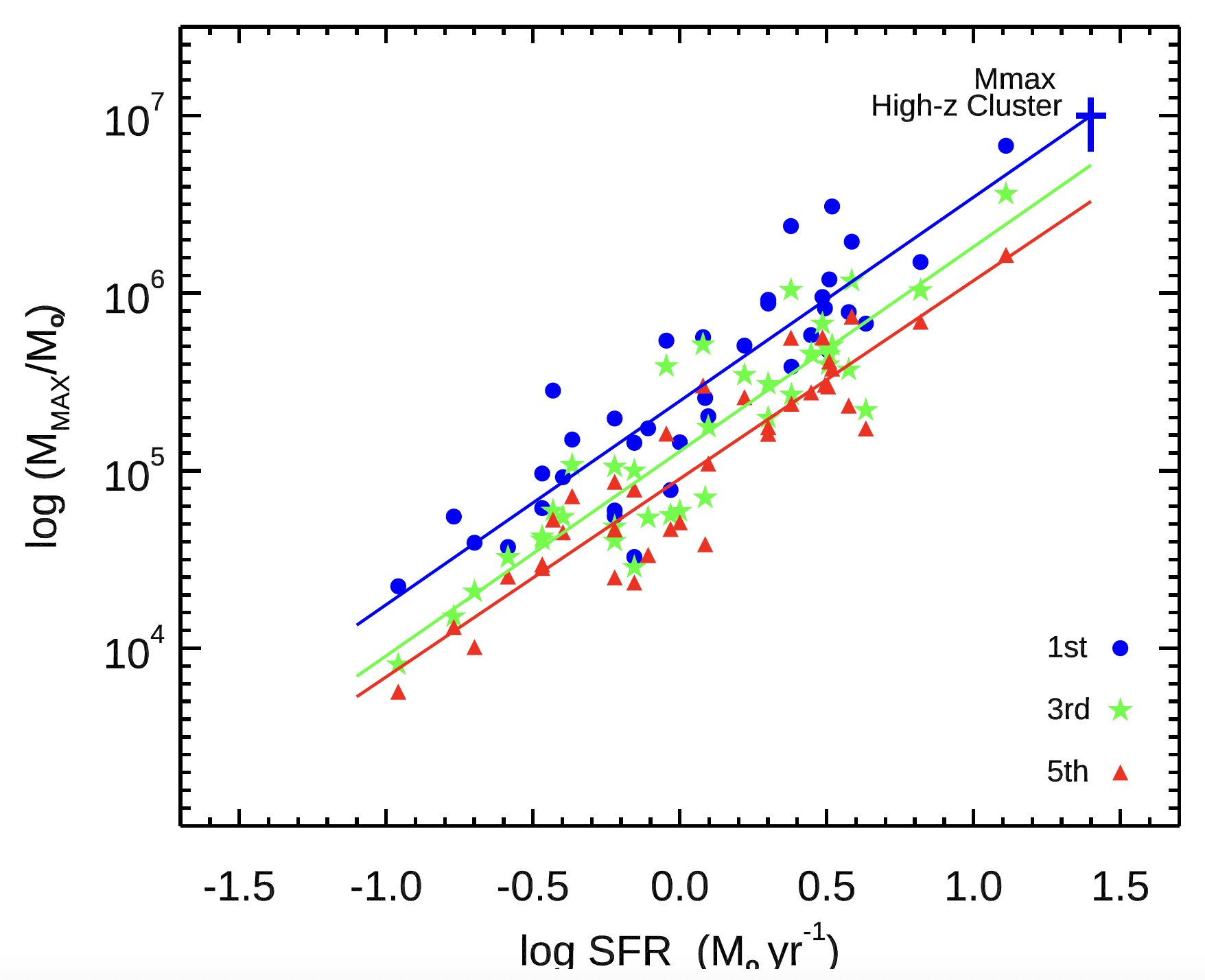} 
	\caption{{\bf The maximum mass of clusters younger than 10~Myr, M$_{\rm max}$ is well-correlated with the star formation rate of the host galaxy}.  The first, third, and fifth most massive clusters are plotted as indicated by the legend. The best fit relations are similar for all three, with slopes of $1.15\pm0.10$ (blue), $1.15\pm0.09$ (green), and $1.12\pm0.08$ (red), but somewhat different y-values.  The cross represents the approximate maximum mass of $10^7~M_{\odot}$ for young clusters discovered in galaxies with redshifts z$>2.5$. The best fit relation for the most massive clusters from the PHANGS sample suggests a star formation of $\approx25~M_{\odot}~\mbox{yr}^{-1}$ for the host galaxy.}\label{fig-Mmaxvssfr}
\end{figure}

In Figure~\ref{fig-Mmaxvssfr} we show that the cluster maximum mass, M$_{\rm max}$, in each of our spiral galaxies is well-correlated with the SFR of the host galaxy.
The figure plots the first, third, and fifth most massive cluster younger than 10~Myr as the blue, green, and red points, respectively, with the associated best fit. For the most massive cluster in a galaxy, we find a best fit of:
\begin{equation}
    \mbox{log}~M_{\rm max}~(M_{\odot}) = 1.15\pm0.01 \times \mbox{log~SFR}~(M_{\odot}~\mbox{yr}^{-1}) + 5.39\pm0.05.
\end{equation}

If we assume that the scaling relationship between cluster masses and SFR of the host galaxy were similar in the early universe to what we see today, we would estimate a SFR for the Sunburst (Sunrise) arc of $25~M_{\odot}~\mbox{yr}^{-1}$ ($9~M_{\odot}~\mbox{yr}^{-1}$) assuming the most massive young cluster has M$_{\rm max}=10^7~M_{\odot}$ ($3\times10^6~M_{\odot}$).
While this estimate and underlying assumptions are speculative, {\em if} high redshift galaxies are shown to follow a similar cluster-galaxy scaling as nearby ones, this would imply that the recently discovered high redshift galaxies are forming stars at rates at the high end but overlapping with star-forming disk environments today.




\section{Conclusions and Future Directions}\label{sec-conclusions}

In this paper, we have studied the mass and age distributions for star cluster populations in 38 nearby, massive spiral galaxies observed as part of the PHANGS collaboration. The sample covers the full range of spiral morphologies from early to late, barred and unbarred, grand-design and flocculent, and spans a factor of $\approx100$ in star formation rate (from $\approx 0.1$ to  $13~M_{\odot}~\mbox{yr}^{-1}$) and surface density of star formation and molecular gas.
Spiral galaxies provide an important benchmark for studies of star cluster populations, since most star formation in today's universe takes place in spiral disks and because star formation has proceeded at a fairly constant rate in them over the past $\approx$Gyr.

The catalogs of compact star clusters are the largest and most uniformly selected in star-forming galaxies to date. Uniform morphological criteria have been applied to select cluster populations in all 38 galaxies, with only single-peaked, centrally concentrated sources selected. Two cluster catalogs have been produced for each galaxy.  Human-based classification has produced state-of-the art catalogs with minimal contamination, which have been supplemented by larger, reproducible machine-learning based catalogs that potentially have a higher degree of contamination.

The PHANGS cluster catalogs include significant improvements to the age-dating methodology which address two of the biggest challenges. For clusters younger than a few 100~Myr, very young, reddened clusters and older clusters with little reddening have the same broadband colors. The catalogs used in this work break this age-reddening degeneracy by using H$\alpha$ emission to identify very young clusters, with a subset of galaxies also using near-infrared measurements in addition to H$\alpha$ emission. For clusters older than $\approx1-2$~Gyr, the age-metallicity degeneracy strongly affects broadband colors.
Candidate globular clusters were identified from their colors and a different set of assumptions adopted for age-dating them.

We have focused on cluster populations on galaxy scales in this work. To further improve statistics and smooth out variations in star formation histories, we also presented cluster mass and age distributions for eight composite groups of 3 to 5 galaxies which have similar star formation rates. We found similar results for the cluster populations in individual galaxies and in composite groups from the human- and machine-learning selected cluster catalogs, and from two methods used for age-dating.  Our main results are as follows.

\begin{itemize}
\item The mass functions of star clusters in spiral galaxies can be described by a power law, $dN/dM \propto M^{\beta}$, with $\beta=-1.9\pm0.2$.  This shape is remarkably similar for the cluster populations in all 38 spirals and 8 galaxy groups which cover a factor of 1000 in cluster mass, from $\approx3\times10^3$ to $\approx 3\times10^6~M_{\odot}$.

\item The power-law index $\beta$ does not vary/correlate with the star formation rate, star formation rate per unit area, or surface density of molecular gas of the host galaxy. The shape of the cluster mass function does not change over the first $\approx$Gyr, but the amplitude decreases continuously over time.

\item The age distributions of star clusters in spiral galaxies decline steeply, and can be described by a power law, $dN/d\tau \propto \tau^{\gamma}$, with $\gamma=-0.83\pm0.23$. This shape is surprisingly similar for the cluster populations in all 38 spirals and 8 galaxy groups.

\item The power-law index $\gamma$ does not correlate with any star-forming property of the host galaxy (i.e.\ SFR, $\Sigma_{\rm SFR}$, or $\Sigma_{\rm H_2}$), or vary with the mass of the clusters.

\item The cluster mass and age distributions in spiral galaxies are independent of one another, $g(M,\tau) \propto M^{-1.9\pm0.2} \tau^{-0.83\pm0.24}$, indicating the populations are dominated by strong, mass-independent dissolution. Dissolution that depends on mass has little impact on cluster populations over $\approx$billion year timescales in spiral disks.

\end{itemize}

These results establish a benchmark for interpreting cluster demographics in different star-forming environments. Future work by the PHANGS team will increase cluster statistics to further reduce uncertainties in observed cluster demographics, since the fine binning in age and mass distributions results in small numbers in some cases, despite starting with large initial samples. The collaboration is currently constructing catalogs of compact star clusters in another 38 spiral galaxies.
We  will also use the extended catalog to examine the relationship between cluster mass and galaxy star formation rate more deeply, by comparing the amplitude of the cluster mass function with SFR.

Our observational results for cluster age-mass demographics in nearby spiral galaxies also provide context for studies of young cluster populations at high redshift.  Some key questions for future studies are to establish whether cluster formation in the early universe was continuous as observed in spiral galaxies or bursty like in merging and post-starburst systems.  The shape of the cluster mass function in early galaxies will directly constrain the role of local environment in cluster formation.

\acknowledgments
The HST data presented in this article were obtained from the Mikulski Archive for Space Telescopes (MAST) at the Space Telescope Science Institute. The specific observations analyzed can be accessed via \dataset[doi: 10.17909/p6nv-br22]{https://doi.org/10.17909/p6nv-br22}.


\dataset[10.17909/t9-r08f-dq31]{https://archive.stsci.edu/hlsp/phangs/phangs-cat} and
\dataset[10.17909/jray-9798]{https://archive.stsci.edu/hlsp/phangs/phangs-hst}.
\bibliography{main,references,references_all}
\bibliographystyle{aasjournal}

\appendix




\section{Results for Individual Galaxies from Primary Catalogs}

\begin{figure}
	\centering
	\includegraphics[width=\linewidth]{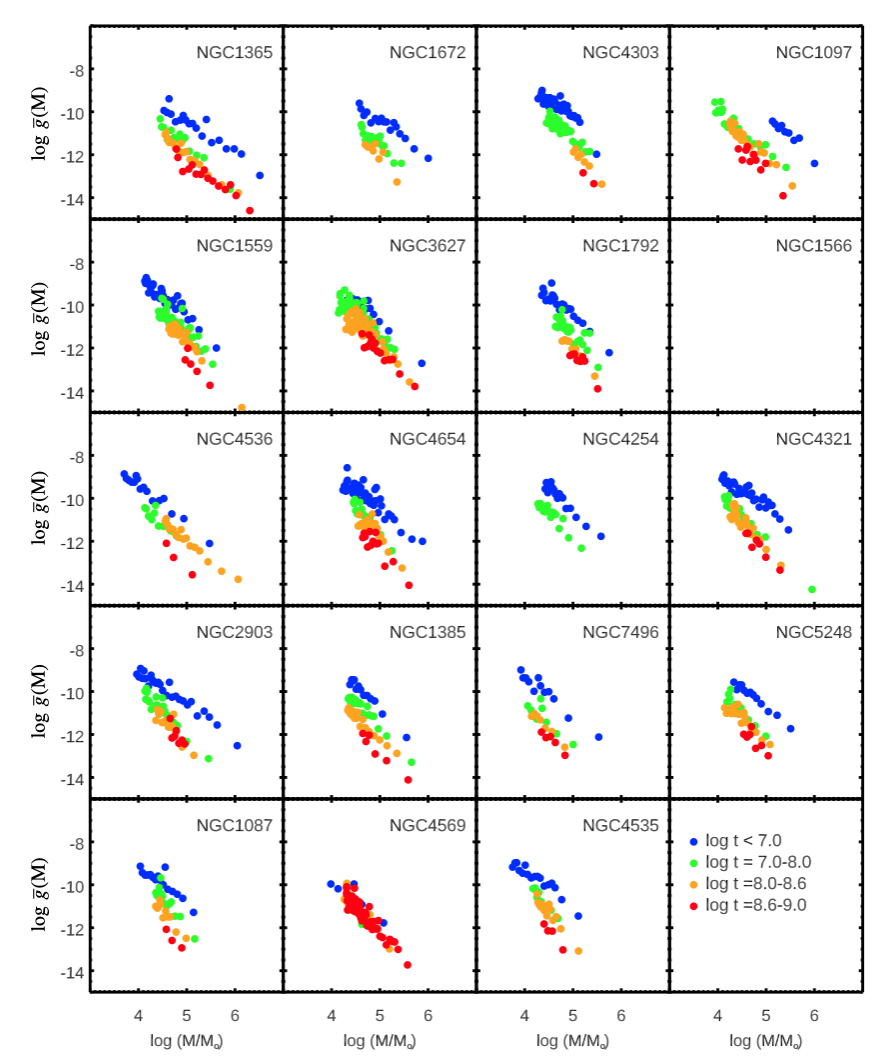}
	\caption{Mass functions of star clusters in 19 PHANGS galaxies
    in the indicated age intervals.  The galaxies are arranged in order of decreasing SFR, and this figure shows the highest 19.
The distributions in each panel are calculated as given in equation~(1) and the relative normalizations are preserved.  The mass functions decrease in amplitude with increasing age in nearly all of the galaxies.}.\label{fig-gM_gals1}
\end{figure}
\begin{figure}
	\centering
	\includegraphics[width=\linewidth]{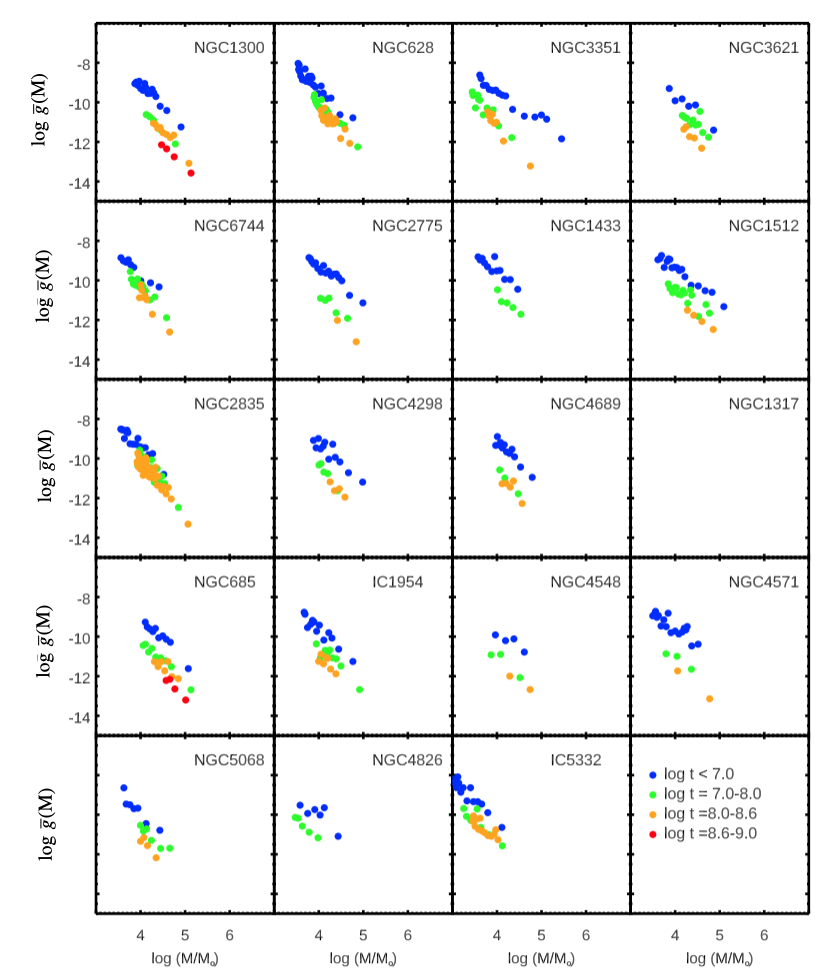}
	\caption{Same as Figure~\ref{fig-gM_gals1} for the 19 galaxies with the lowest rates of star formation in our sample. }.\label{fig-gM_gals2}
\end{figure}

In this Appendix we present the fitting results for the cluster mass and age distributions in all 38 galaxies based on our primary dataset: the human-classified catalogs with uniform age-dating from \citet{Thilker25}. Figures~\ref{fig-gM_gals1} and \ref{fig-gM_gals2} show the cluster mass distributions, starting with NGC~1365 (upper left panel) which has the highest rate of star formation, in decreasing order.  The figures use the same age intervals and color-coding as Figure~\ref{fig-gM_groups} for the galaxy groups: (1) log~($\tau/\mbox{yr})<7.0$ (blue), (2) log~($\tau/\mbox{yr})=7.0-8.0$ (green), (3) log~($\tau/\mbox{yr}) = 8.0-8.6$ (orange) and (4) log~($\tau/\mbox{yr})=8.6-9.0$ (red). 
There are an insufficient number of clusters with ages between log~($\tau/\mbox{yr})=8.6-9.0$ to construct the $g(M)$ distribution in many galaxies. The best-fit power-law index $\beta$ is recorded in Table~\ref{tab:betagals} along with the lower mass limit used for the fit.  



Overall, the results for the individual galaxies are similar to those found for the cluster populations in the galaxy groups. To first order, the shapes of the cluster mass functions are similar in different age intervals and different galaxies. For the youngest clusters (log~$\tau/\mbox{yr})=6.0-7.0$) across the entire sample, the median power-law index is $\beta=-1.83$, with a mean and standard deviation of $-1.86\pm0.24$. The median (mean) values of $\beta$ for next three age intervals are somewhat steeper (at the $1\sigma$ level), with a mean and standard deviation of $-2.11\pm0.31$ (log~($\tau/\mbox{yr})=7.0-8.0$), $-2.23\pm0.32$ (log~($\tau/\mbox{yr})=8.0-8.6$), and $-2.24\pm0.26$ (log~($\tau/\mbox{yr})=8.6-9.6$), respectively. These results indicate that the shape of the cluster mass function does not evolve (flatten) over the first Gyr. 

In almost every single galaxy, the amplitude of the cluster mass function decreases with increasing age, with a declining sequence from blue to green to orange to red. There are however a handful of cases where the amplitudes of the mass functions are nearly identical for two (or more) age intervals. The best example of this is NGC~4569, which likely experienced a higher SFR in the past than at present.

\begin{deluxetable*}{lcccccccccc}
\tablecaption{Power-law Index $\beta$ for Mass Functions in Individual Galaxies From Primary Catalogs}
\centering
\tablehead{
\colhead{Galaxy}  & \colhead{$\beta$1} &  \colhead{M$_{\rm lim1}$} & \colhead{$\beta$2} & \colhead{M$_{\rm lim2}$} & \colhead{$\beta$3} & \colhead{M$_{\rm lim3}$} & \colhead{$\beta$4} & \colhead{M$_{\rm lim4}$}  \\ 
 \colhead{Name} & \colhead{(1-10~Myr)} & \colhead{logM} & \colhead{(10-100~Myr)} & \colhead{logM} & \colhead{(100-400~Myr)} & \colhead{logM} &
 \colhead{(400~Myr-1~Gyr)} & \colhead{logM}  
 }
\startdata
IC1954 & $-2.04\pm0.21$ & 3.6 &  $-2.11\pm0.31$ & 3.9 & $-1.98\pm0.52$ & 3.9  & ... & ... \\
IC~5332 & $-2.10\pm0.21$ & 3.0 & $-2.03\pm0.39$ & 3.2 &  $-1.58\pm0.34$ & 3.4  & ... & ...  \\
NGC~628 & $-2.14\pm0.15$ & 3.5 & $-2.32\pm0.18$ & 3.8 & $-2.09\pm0.36$ & 4.0 & ... & ...  \\
NGC~685 & $-2.12\pm0.19$ & 4.0 & $-1.99\pm0.14$ & 4.0 & $-1.56\pm0.55$ & 4.2 & $-2.42\pm0.41$ &  4.2  \\
NGC~1087 & $-1.64\pm0.21$ & 4.0 & $-2.49\pm0.32$ & 4.3 & $-2.71\pm0.36$ & 4.3 & $-2.50\pm0.69$ & 4.4  \\
NCC~1097 & $-2.00\pm0.21$ & 5.0 &  $-1.97\pm0.11$ & 3.9 & $-2.05\pm0.21$ & 4.2 & $-2.11\pm0.38$ & 4.3  \\
NGC~1300 & $-2.11\pm0.17$ & 3.8 & $-2.30\pm0.11$ & 4.0 &  $-2.27\pm0.25$ & 4.2 & ... & ... \\
NGC~1317 & ... & ... & ... & ... & ... & ... & ... & ...  \\
NGC~1365 & $-1.56\pm0.15$ & 5.0 & $-1.81\pm0.10$ & 4.2 & $-1.95\pm0.11$ & 4.4 & $-1.50\pm0.11$ & 4.7  \\
NGC~1385 & $-2.22\pm0.11$ & 4.3 & $-2.25\pm0.12$ & 4.3 & $-2.08\pm0.12$ & 4.3 & $-2.22\pm0.29$ & 4.5  \\
NGC~1433 & $-1.68\pm0.25$ & 3.5 & $-2.05\pm0.41$ & 3.9 & ... & ... & ... & ...  \\
NGC~1512 & $-1.66\pm0.10$ & 3.5 & $-1.43\pm0.26$ & 3.8 & $-1.66\pm0.04$ & 4.0 & ... & ...  \\
NGC~1559 & $-1.83\pm0.11$ & 4.1 & $-1.96\pm0.17$ & 4.4 & $-2.26\pm0.12$ & 4.6 & $-2.40\pm0.32$ & 4.7  \\
NGC~1566 & $-1.79\pm0.10$ & 4.1 & $-2.21\pm0.26$ & 4.4 & $-2.15\pm0.19$ & 4.4 & $-2.16\pm0.51$ & 4.8   \\
NGC~1672 & $-1.95\pm0.19$ & 5.0 & $-2.02\pm0.20$ & 4.5 & $-2.45\pm0.38$ & 4.65  & ... & ... \\
NGC~1792 & $-1.93\pm0.14$ & 4.3 & $-2.00\pm0.27$ & 4.6 & $-2.51\pm0.23$ & 4.7 & $-2.55\pm0.51$ & 4.8  \\
NGC~2775 & $-1.74\pm0.12$ & 3.7 & $-1.85\pm0.41$ & 3.9 & ... & ... & ... & ...  \\
NGC~2835 & $-2.06\pm0.18$ & 3.5 & $-2.51\pm0.31$ & 3.9 &  $-2.44\pm0.17$ & 3.9 & ... & ...  \\
NGC~2903 & $-1.51\pm0.07$ & 3.9 & $-2.30\pm0.17$ & 4.1 & $-2.54\pm0.31$ & 4.3 & ... & ...  \\
NGC~3351 & $-1.49\pm0.10$ & 3.4 & $-2.38\pm0.31$ & 3.4 & $-2.60\pm0.29$ & 3.7  & ... & ...  \\
NGC~3621 & $-1.80\pm0.29$ & 3.5 & $-1.42\pm0.54$ & 4.1 & $-2.55\pm0.45$ & 4.1 & ... & ... \\
NGC~3627 & $-2.02\pm0.13$ & 4.3 & $-1.95\pm0.13$ & 4.1 & $-2.04\pm0.15$ & 4.1 & $-2.08\pm0.20$ &  4.5  \\
NGC~4254 & $-2.09\pm0.14$ & 4.3 & $-2.23\pm0.22$ & 4.3 & $-2.36\pm0.55$ & 4.6  &  ... & ...  \\
NGC~4298 & $-1.92\pm0.26$ &  3.7 & $-2.02\pm0.39$ & 4.1 & $-1.99\pm0.67$ & 4.1 & ... & ...  \\
NGC~4303 & $-1.66\pm0.16$ & 4.2 & $-1.81\pm0.16$ & 4.4 & $-1.91\pm0.31$ & 4.7 & ... & ...  \\
NGC~4321 & $-2.05\pm0.36$ & 4.7 & $-2.09\pm0.10$ & 4.1 & $-2.44\pm0.21$ & 4.2 & $-2.39\pm0.35$ & 4.5  \\
NGC~4535 & $-1.59\pm0.13$ & 3.6 & $-2.77\pm0.31$ & 4.1 & $-2.32\pm0.18$  & 4.1 & ... & ...  \\
NGC~4536 & $-1.74\pm0.10$ & 3.6 & $-1.88\pm0.36$ & 4.1 & $-2.08\pm0.25$ & 4.1 & $-2.50\pm0.52$ & 4.3  \\
NGC~4548 & ... & ... & ... & ... & ... & ... & ... & ...  \\
NGC~4569 & $-1.98\pm0.75$ & 4.0 & ... & ... & $-2.46\pm0.19$ & 4.2 & $-2.05\pm0.11$ & 4.3  \\
NGC~4571 & $-1.38\pm0.20$ & 3.4 & $-1.76\pm0.30$  & 3.5 & $-1.72\pm0.30$ &  3.5 & ... & ...  \\
NGC~4654 & $-1.76\pm0.12$ & 4.3 & $-2.67\pm0.25$ & 4.4 & $-2.18\pm0.29$  & 4.5 & $-2.24\pm0.34$ & 4.5   \\
NGC~4689 & $-2.23\pm0.23$ & 3.9 & $-2.76\pm0.21$ & 3.9 & $-2.05\pm0.71$ & 4.0 & ... & ...  \\
NGC~4826 & $-1.47\pm0.56$ & 3.4 & $-2.11\pm0.24$ & 3.4 & ... & ... & ... & ...  \\
NGC~5068 & $-2.29\pm0.39$ & 3.5 & $-1.87\pm0.36$ & 3.9 & $-2.69\pm0.74$ & 3.9 & ... & ...  \\
NGC~5248 & $-1.88\pm0.09$ & 3.9 & $-2.38\pm0.23$ & 4.1 & $-1.92\pm0.21$ & 4.1 & $-2.11\pm0.65$ & 4.2  \\
NGC~6744 & $-1.91\pm0.20$ & 3.5 &  $-2.34\pm0.20$ & 3.7 & $-3.00\pm0.47$ & 3.9 & ... & ...  \\
NGC~7496 & $-1.83\pm0.15$ & 3.8 & $-1.99\pm0.48$ & 4.0 &  $-2.24\pm0.19$ & 4.0 & $-2.08\pm0.33$ & 4.2  \\
\enddata
\tablecomments{The power-law index $\beta$ measured for the given age interval in each galaxy from our primary catalogs.  The lower-mass limit for the distribution is listed after each age interval.}
\label{tab:betagals}
\end{deluxetable*}

\begin{figure}
	\centering
\includegraphics[width=18.0cm]{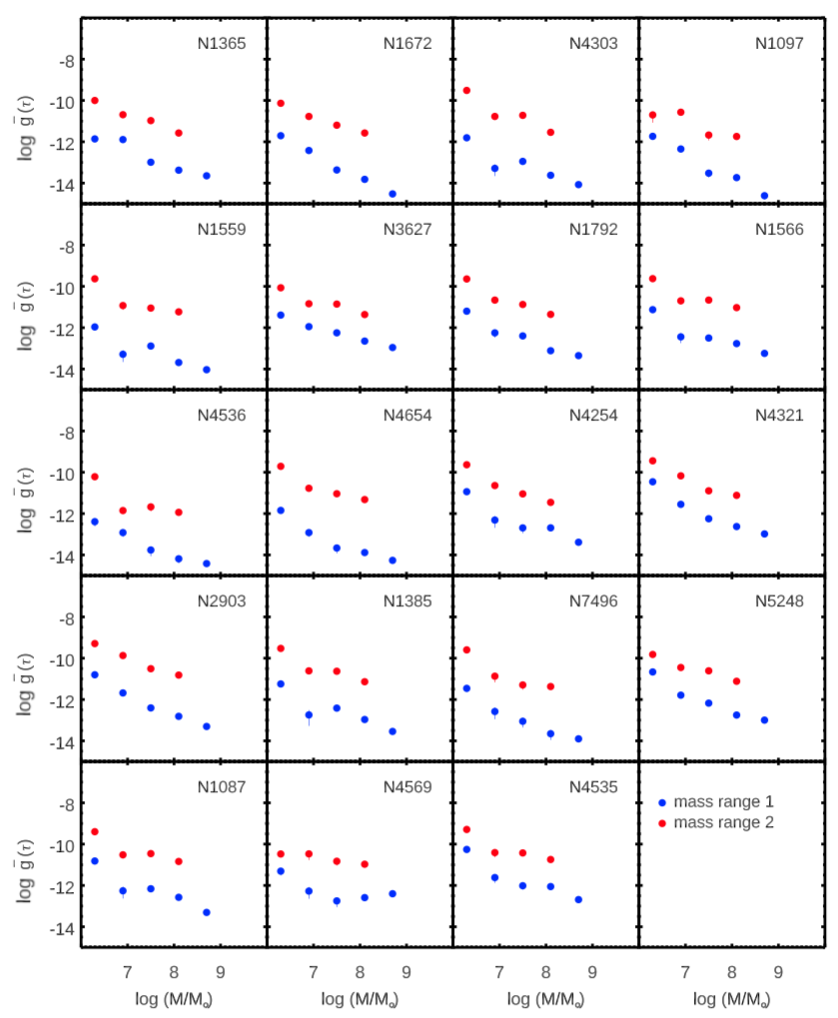}
	\caption{Age distributions $g(\tau)$ for the 19 galaxies with the highest SFR in our sample.  Distributions are shown for 2 different intervals of mass (listed in Table~\ref{tab:gammagals}).}.
    \label{fig-gt_gals}
\end{figure}

\begin{figure}
	\centering
\includegraphics[width=18.0cm]{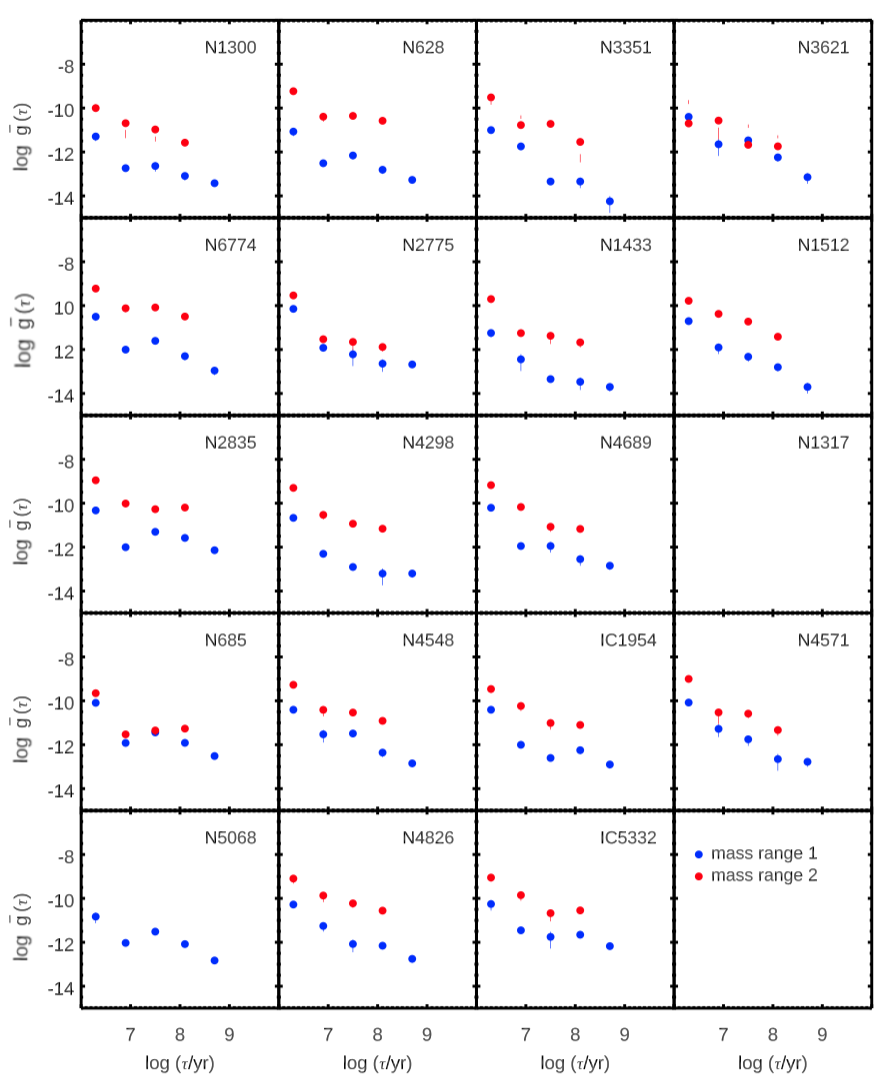}
	\caption{Same as Figure~\ref{fig-gt_gals} but for the second half of the sample in terms of SFR.}
    \label{fig-gt_gals2}
\end{figure}

In Figures~\ref{fig-gt_gals} and \ref{fig-gt_gals2} we plot the cluster age distributions $g(\tau)$ for two adjacent mass intervals in each galaxy.  
The distributions were created using 
the same fixed age bins used for the galaxy groups.
The distributions are noisier than for the groups, but have the same declining shape and can be described by a power-law.
The best-fit power-law index $\gamma$ is recorded in Table~\ref{tab:gammagals}, along with the mass and age intervals used in each fit.
The $\gamma$ values are nearly all negative and range from $\sim -0.5$ to $\sim -1.5$, albeit with large uncertainties in a number of cases.  

The fits to individual galaxies support the results found for the composite groups, that the age distributions of star clusters in spiral galaxies decline steeply and more-or-less continuously over the past $\approx1$~Gyr, and that this decline does not depend on the masses of the clusters.
Across the entire spiral sample, the most massive clusters have a median (mean$+$ stddev) of $\gamma=-0.88$ ($-0.90\pm0.18$) when fitting from 1~Myr to 1~Gyr, and $-0.75$ ($-0.73\pm0.31$) when the fits are restricted to clusters with ages between 15~Myr and 1~Gyr, which should all be bound.  The age distribution for the lower-mass interval has $-0.83$ ($-0.85\pm0.20$), so the shape of the age distribution does not change with cluster mass.
The results for $\gamma$ are plotted as the small open circles in Figure~\ref{fig-dndt-gamma}.


\begin{deluxetable*}{lccccccccc}
\tablecaption{Power-law Index for Age Distribution Fits to Individual Galaxies From Primary Catalogs}
\centering
\tablehead{
\colhead{Galaxy} & \colhead{SFR} & \multicolumn{3}{c}{Mass Range 1} & \multicolumn{2}{c}{Mass Range 2}   \\ 
 \colhead{\#} & \colhead{$M_{\odot}~\mbox{yr}^{-1}$} & \colhead{$\gamma_{\rm 1~Myr-1~Gyr}$} & \colhead{$\gamma_{\rm 15~Myr-1~Gyr}$} & \colhead{logM} &
 \colhead{$\gamma_{\rm 1-250~Myr}$} & \colhead{logM} 
 }
\startdata
NGC~1365 & 12.7 & $-0.84\pm0.14$ & $-0.55\pm0.05$ & $>5.0$ & $-0.83\pm0.09$ & $4.5-5.0$ \\
NGC~1672 & 6.6 & $-1.17\pm0.07$ & $-0.97\pm0.12$ & $>5.0$ & $-0.79\pm0.07$ & $4.7-5.0$ \\
NGC~4303 & 4.3 & $-0.81\pm0.23$ & $-0.94\pm0.11$ & $>5.0$ & $-1.00\pm0.28$ & $4.7-5.0$ \\
NGC~1097 & 3.85 & $-1.19\pm0.12$ & $-0.91\pm0.32$ & $>5.0$ & $-0.70\pm0.27$ & $4.7-5.0$ \\
NGC~1559 & 3.76 & $-0.76\pm0.21$ & $-0.95\pm0.22$ & $>5.0$ & $-0.82\pm0.33$ & $4.7-5.0$ \\
NGC~3627 & 3.3 & $-0.64\pm0.04$ & $-0.59\pm0.04$ & $>4.8$ & $-0.65\pm0.16$ & $4.5-4.8$ \\
NGC~1792 & 3.23 & $-0.86\pm0.13$ & $-0.79\pm0.23$ & $>5.0$ & $-0.89\pm0.19$ & $4.5-5.0$ \\
NGC~1566 & 3.2 & $-0.76\pm0.19$ & $-0.62\pm0.10$ & $>4.8$ & $-0.69\pm0.26$ & $4.4-4.8$ \\
NGC~4536 & 3.12 & $-0.89\pm0.10$ & $-0.55\pm0.09$ & $>4.7$ & $-0.83\pm0.44$ & $4.4-4.7$ \\
NGC~4654 & 3.06 & $-0.97\pm0.17$ & $-0.50\pm0.08$ & $>5.0$ & $-0.85\pm0.22$ & $4.6-5.0$ \\
 NGC~4254 & 2.8 & $-0.88\pm0.22$  & $-0.58\pm0.34$ & $>4.9$ & $-0.98\pm0.17$  & $4.4-4.9$ \\
NGC~4321 & 2.4 & $-1.02\pm0.15$  & $-0.61\pm0.01$ & $>4.7$ & $-0.95\pm0.15$ & $4.4-4.7$ \\
NGC~2903 & 2.39 & $-1.02\pm0.09$  & $-0.75\pm0.04$  & $>4.5$  & $-0.87\pm0.08$  & $4.2-4.5$ \\
NGC~1385 & 2.0 & $-0.80\pm0.23$  & $-0.94\pm0.01$ & $>4.8$ & $-0.81\pm0.24$ & $4.4-4.8$ \\
NGC~7496 & 2.0 & $-0.99\pm0.14$  & $-0.71\pm0.17$ & $>4.6$ & $-0.96\pm0.31$ & $4.2-4.6$ \\
NGC~5248 & 1.66 & $-0.94\pm0.14$  & $-0.68\pm0.16$  & $>4.5$ & $-0.67\pm0.10$ & $4.2-4.5$ \\
 NGC~1087 & 1.25 & $-0.88\pm0.21$  & $-0.95\pm0.15$ & $>4.6$ & $-0.71\pm0.27$ & $4.2-4.6$ \\
 NGC4569 & 1.22 & $-0.42\pm0.25$  & $+0.29\pm0.02$ & $>4.4$ & $-0.31\pm 0.09$ & $4.1-4.4$ \\
 NGC~4535 & 1.2 & $-0.88\pm0.21$  & $-0.56\pm0.28$ & $>4.4$ & $-0.73\pm0.27$ & $4.1-4.4$ \\
NGC~1300 & 1.0 &  $-0.77\pm0.22$ & $-0.65\pm0.06$ & $>4.6$ & $-0.91\pm0.25$ & $4.3-4.6$ \\
 NGC~628 & 0.93 & $-0.78\pm0.23$  & $-0.93\pm0.10$ & $>4.5$ & $-0.67\pm0.30$ & $4.0-4.5$ \\
 NGC~3351 & 0.9 & $-1.34\pm0.21$  & $-0.75\pm0.43$ & $>4.5$  & $-1.47\pm0.32$ & $4,0-4.5$ \\
 NGC~3621 & 0.78 & $-1.01\pm0.19$  & $-1.39\pm0.06$  & $>4.4$ & $-0.78\pm0.31$ & $4.1-4.4$ \\
  NGC~6744 & 0.71 & $-0.87\pm0.24$  & $-1.13\pm0.02$ & $>4.2$ & $-0.63\pm0.21$ & $3.7-4.2$ \\
 NGC~2775 & 0.70 & $-0.96\pm0.30$  & $-0.65\pm0.09$  & $>4.5$ & $-1.19\pm0.52$ & $4.2-4.5$ \\
 NGC~1433 & 0.6 & $-0.99\pm0.23$  & $-0.30\pm0.50$ & $>4.5$ & $-1.00\pm0.38$ & $4.2-4.5$ \\
 NGC~1512 & 0.6 & $-1.15\pm0.12$  & $-1.15\pm0.21$ & $>4.5$ & $-0.89\pm0.08$ & $4.1-4.5$ \\
 NGC~2835 & 0.6 & $-0.53\pm0.31$  & $-0.70\pm0.14$ & $>4.2$ & $-0.66\pm0.30$ & $3.8-4.2$ \\
 NGC~4298 & 0.43 &  $-0.99\pm0.30$ & $-0.25\pm0.15$ & $>4.5$ & $-0.99\pm0.27$ & $4.0-4.5$ \\
 NGC~4689 & 0.4 & $-0.98\pm0.26$  & $-0.75\pm0.14$ & $>4.4$ & $-1.15\pm0.25$ & $4.0-4.4$ \\
 NGC~1317 & ... & ... & ... & ... & ... & ... \\
 NGC~685 & 0.37 &  $-0.81\pm0.30$ & $-0.90\pm0.06$ & $>4.5$ & $-0.78\pm0.57$ & $4.2-4.5$ \\
  NGC~4548 & 0.34 & $-0.87\pm 0.31$  & $-0.25\pm0.48$ & $>4.2$ & $-0.95\pm0.20$ & $3.8-4.2$ \\
 IC~1954 & 0.34 & $-0.95\pm0.14$  & $-1.13\pm0.18$ & $>4.2$ & $-0.84\pm0.25$ & $3.8-4.2$ \\
NGC~4571 & 0.26 & $-1.12\pm0.16$  & $-0.85\pm0.37$ & $>4.2$ & $-1.17\pm0.32$  & $3.8-4.2$ \\
NGC~5068 & 0.20 & $-0.67\pm0.23$  & $-1.09\pm0.09$ & $>4.0$ & $-0.80\pm0.25$ & $3.7-4.0$ \\
 NGC~4826 & 0.17 & $-0.97\pm0.15$  & $-0.56\pm0.25$  & $>3.6$  & $-0.79\pm0.12$  & $3.2-3.6$ \\
 IC~5332 & 0.11 & $-0.67\pm0.20$  & $-0.88\pm0.20$  & $>3.8$  & $-0.88\pm0.27$  & $3.5-3.8$ \\
 \enddata
\tablecomments{The power-law index $\gamma$ for the age distribution measured in the listed mass intervals for each galaxy. }
\label{tab:gammagals}
\end{deluxetable*}

\section{Results for Individual Galaxies from Secondary Catalogs}


Here we use the improved age and mass estimates for the cluster catalogs in 16 galaxies based on 10-filter SED fitting presented in \citet{Henny25}. These catalogs contain between 50\% and 95\% of the full cluster lists in the primary catalogs, and are presented as a point of comparison based on an independent age-dating methodology.

\begin{figure}[ht]
	\centering
	\includegraphics[width=18.5cm]{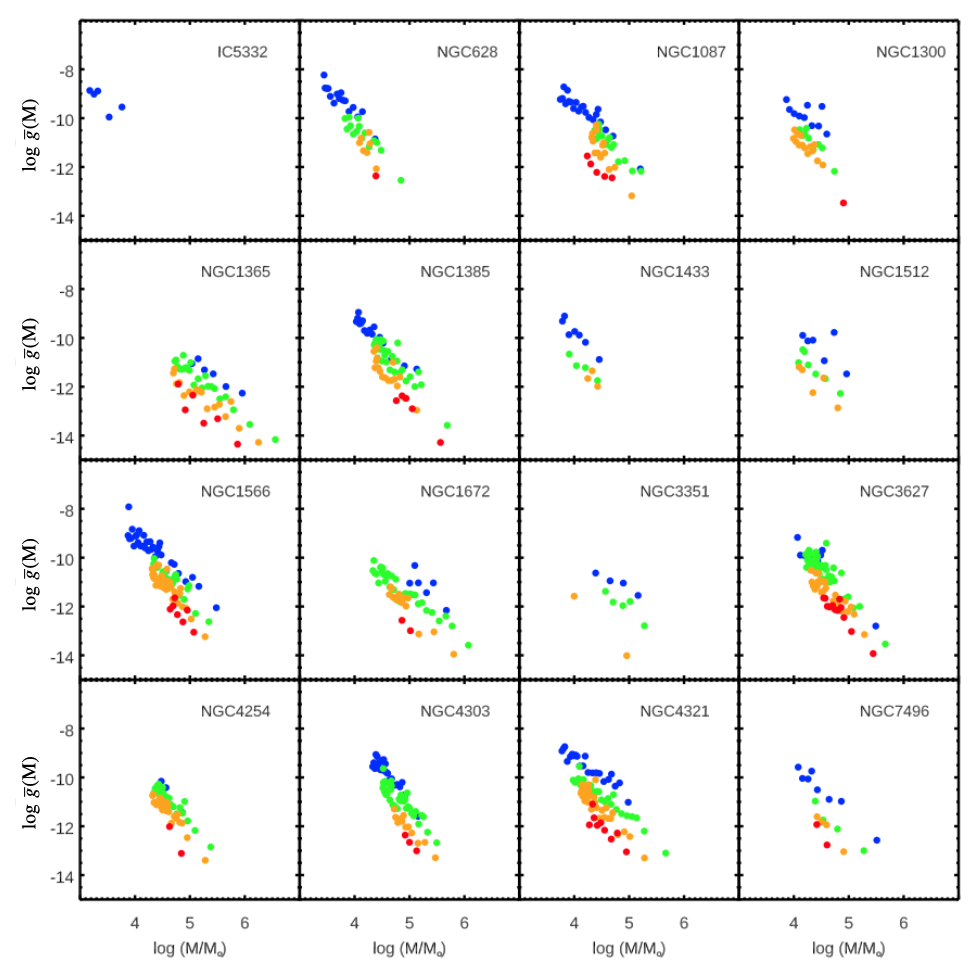}
	\caption{Mass function of star clusters constructed from the secondary catalogs of \citet{Henny25} for 16 galaxies}\label{fig-gMKiana}
\end{figure}

Figure~\ref{fig-gMKiana} presents the mass functions and Table~\ref{tab:betahenny} records the best-fit values of $\beta$ for the four age intervals used throughout this work. $\beta$ for $1-10$~Myr clusters over the 16 spiral galaxies in the \citet{Henny25} catalogs has a median (mean$+$stddev) of $-1.9$ ($-1.7\pm0.3$). The mass function of $10-100$~Myr clusters is somewhat steeper, at the $\approx1\sigma$ level, with median (mean) $\beta$ values of $-2.12$ ($-2.04 \pm0.24$). The amplitudes of the cluster mass function in every galaxy decreases for older clusters. Overall, the shape and amplitude of the cluster mass function for this subset with improved age-dating is consistent with those from the primary catalogs.

\begin{figure}[ht]
	\centering
	\includegraphics[width=\textwidth]{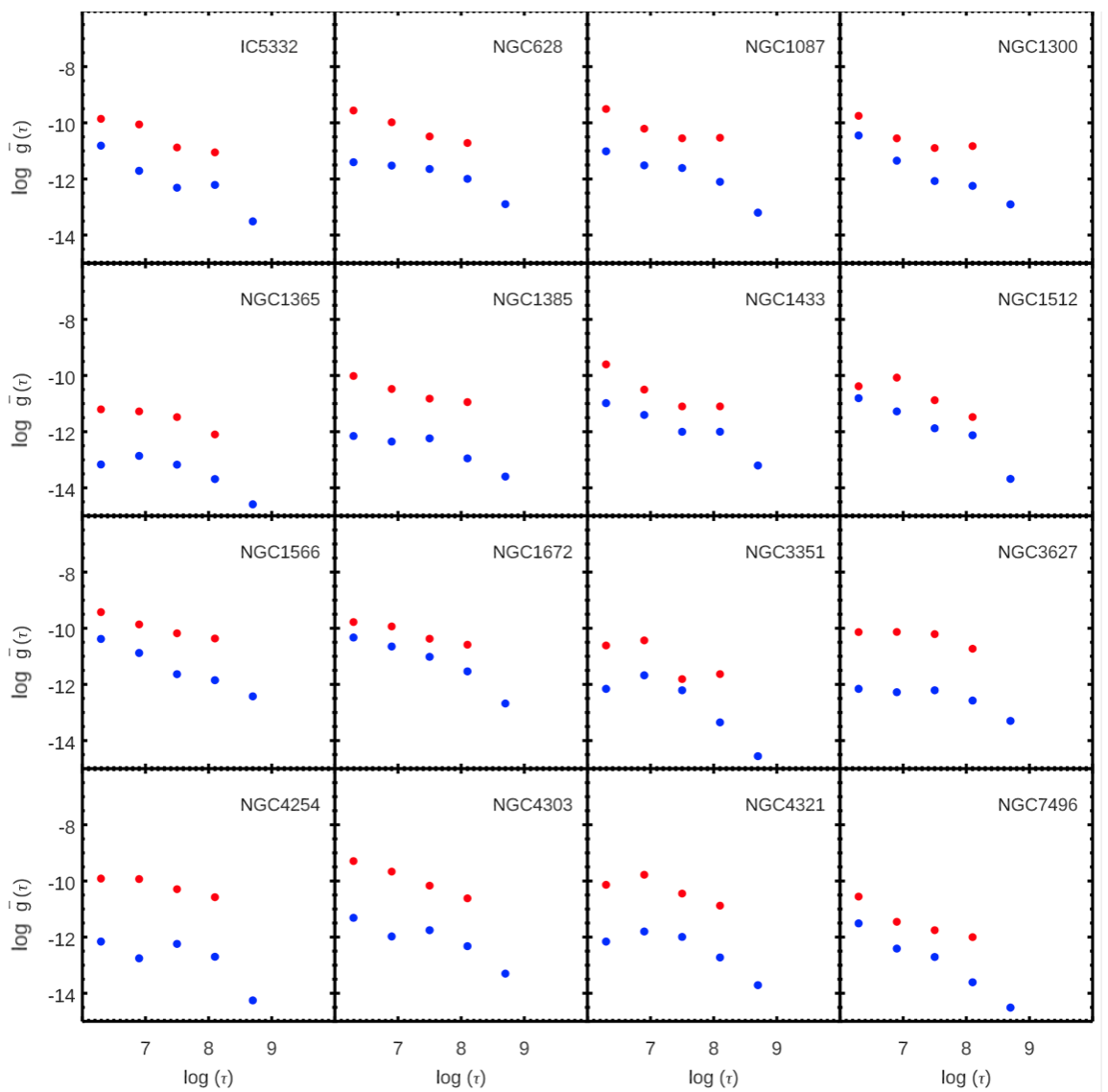}
	\caption{Age distributions of star clusters constructed from the secondary catalogs of \citet{Henny25} for 16 galaxies.}\label{fig-gtKiana}
\end{figure}

Age distributions in two adjacent mass intervals are presented in Figure~\ref{fig-gtKiana} and the best-fit power-law indices are compiled in Table~\ref{tab:gammahenny}. We find a median of $-0.82$ and mean $-0.82\pm0.22$ for the higher mass interval, $-0.57$ and $-0.60\pm0.15$ for the lower-mass interval, which are within $1\sigma$.  The age distributions from the secondary catalogs are consistent with the results from the primary catalogs, and do not show statistically significant differences in the shape of the age distribution with cluster mass.

\begin{deluxetable*}{lccccccccc}
\tablecaption{Power-law Index $\beta$ for Mass Function Fits to Secondary Catalogs\label{tab:betahenny}}
\centering
\tablehead{
\colhead{Galaxy} & \colhead{$\beta1$} & \colhead{$\beta2$} & \colhead{$\beta3$} & \colhead{$\beta4$}  & \colhead{\#} \\
 \colhead{Name} & \colhead{1-10~Myr}  & \colhead{10-100~Myr} & \colhead{100-400~Myr} & \colhead{1Myr-1~Gyr} & \colhead{Clusters}
 }
\startdata
IC5332 & $-1.96\pm0.21$ & ... & ... & ... & 166 \\
NGC~628 & $-2.09\pm0.22$ & $-2.14\pm0.26$ & $-2.57\pm1.12$ & ... & 320 \\
NGC~1087 & $-2.11\pm0.22$ & $-2.05\pm0.21$ & ... & ... & 482  \\
NGC~1300 & $-1.80\pm0.92$ & $-2.52\pm0.57$ & ... & ... & 257 \\
NGC~1365 & $-1.52\pm0.20$ & $-1.80\pm0.10$ & $-1.58\pm0.16$ & $-1.92\pm0.46$ & 557 \\
NGC~1385 & $-2.09\pm0.15$ & $-2.18\pm0.21$ & $-2.53\pm0.38$ & $-2.26\pm0.35$ & 503 \\
NGC~1433 & $-2.32\pm0.33$ & $-1.92\pm0.33$ & $-2.00\pm1.20$ & ... & 173 \\
NGC~1512 & $-1.48\pm0.34$ & $-2.14\pm0.41$ & $-2.02\pm0.71$ & ... & 220 \\
NGC~1566 & $-1.86\pm0.13$ & $-2.12\pm0.20$ & $-2.49\pm0.22$ & $-2.25\pm0.84$ & 622 \\
NGC~1672 & $-1.90\pm0.65$ & $-1.72\pm0.09$ & $-2.26\pm0.20$ & ... & 353 \\
NGC~3351 & $-1.24\pm0.14$ & $-1.66\pm0.46$ & ... & ... & 162 \\
NGC~3627 & $-2.06\pm0.27$ & $-2.24\pm0.18$ & $-2.13\pm0.29$ & $-2.41\pm0.29$ & 551 \\
NGC~4254 & ... & $-2.27\pm0.16$ & $-2.43\pm0.18$ & ... & 411 \\
NGC~4303 & $-2.50\pm0.22$ & $-2.16\pm0.18$ & $-2.40\pm0.22$ & ... & 512 \\
NGC~4321 & $-1.53\pm0.12$ & $-1.71\pm0.09$ & $-2.35\pm0.19$ & $-2.07\pm0.47$ & 601 \\
NGC~7496 & $-1.88\pm0.18$ & $-2.03\pm0.36$ & ... & ... & 236 \\
\enddata
\end{deluxetable*}

\begin{deluxetable*}{lccccccccc}
\tablecaption{Power-law Index for Age Distribution Fits to Secondary Catalogs\label{tab:gammahenny}}
\centering
\tablehead{
\colhead{Group} & \multicolumn{3}{c}{Mass Range 1} & \multicolumn{2}{c}{Mass Range 2}    \\ 
 \colhead{\#} & \colhead{$\gamma_1$ (1~Myr-1~Gyr)} & \colhead{$\gamma_2$ (15~Myr-1~Gyr)} & \colhead{logM} &
 \colhead{$\gamma_3$ (1-250~Myr)} & \colhead{logM} 
 }
\startdata
IC5332 & $-0.98\pm0.19$ & $-1.00\pm0.67$ & $>3.8$ & $-0.73\pm0.15$ & $3.4-3.8$ \\
NGC~628 & $-0.58\pm0.15$ & $-1.04\pm0.27$ & $>4.2$ & $-0.66\pm0.06$ & $3.8-4.3$ \\
NGC~1087 & $-0.82\pm0.16$ & $-1.32\pm0.29$ & $>4.4$ & $-0.57\pm0.19$ & $4.0-4.4$ \\
NGC~1300 & $-0.97\pm0.12$ & $-0.69\pm0.23$ & $>4.4$ & $-0.59\pm0.23$ & $4.0-4.4$ \\
NGC~1365 & $-0.61\pm0.22$ & $-1.18\pm0.19$ & $>5.0$ & $-0.48\pm0.15$ & $4.6-5.0$ \\
NGC~1385 & $-0.58\pm0.16$ & $-1.13\pm0.03$ & $>4.7$ & $-0.52\pm0.09$ & $4.2-4.7$ \\
NGC~1433 & $-0.84\pm0.16$ & $-1.00\pm0.58$ & $>4.2$ & $-0.85\pm0.24$ & $3.9-4.2$ \\
NGC~1512 & $-1.09\pm0.20$ &  $-1.50\pm0.63$ & $>4.5$  & $-0.68\pm0.28$  & $4.2-4.5$  \\
NGC~1566 & $-0.84\pm0.07$  &  $-0.66\pm0.17$  & $>4.8$  & $-0.53\pm0.08$  & $4.4-4.8$  \\
NGC~1672 & $-0.93\pm0.16$  &  $-1.38\pm0.32$  & $>4.8$  & $-0.48\pm0.06$  & $4.5-4.8$  \\
NGC~3351 & $-1.08\pm0.34$  &  $-1.95\pm0.05$  & $>4.3$  & $-0.74\pm0.36$  & $3.8-4.3$  \\
NGC~3627 & $-0.44\pm0.15$  &  $-0.91\pm0.17$  & $>4.9$  & $-0.34\pm0.15$  & $4.4-4.9$  \\
NGC~4254 & $-0.69\pm0.32$  &  $-1.50\pm0.52$  & $>4.7$  & $-0.39\pm0.09$  & $4.3-4.7$  \\
NGC~4303 & $-0.72\pm0.19$  &  $-1.28\pm0.23$  & $>4.8$  & $-0.74\pm0.03$  & $4.4-4.8$  \\
NGC~4321 & $-0.67\pm0.26$  &  $-1.43\pm0.20$  & $>4.8$  & $-0.51\pm0.26$  & $4.4-4.8$  \\
NGC~7496 &  $-1.20\pm0.12$  &  $-1.50\pm0.01$ & $>4.8$  & $-0.77\pm0.18$  & $4.4-4.8$  \\
\enddata
\end{deluxetable*}

\end{document}